\documentclass[11pt]{article}
\usepackage[dvipsnames]{xcolor}
\usepackage{amsmath, amssymb, amsfonts, setspace, url, hyperref, verbatim, cite, graphicx, todonotes, cancel, slashed, mathtools,amsthm}
\usepackage{amsthm}
\usepackage{makecell}

\usepackage[bbgreekl]{mathbbol}
\usepackage{amssymb}
\usepackage{bbm}
\DeclareSymbolFontAlphabet{\mathbb}{AMSb}
\DeclareSymbolFontAlphabet{\mathbbalternative}{bbold}

\usepackage{ stmaryrd }
\usepackage{dsfont,float}
\usepackage{bbm}
\usepackage[T1]{fontenc}
\usepackage[utf8]{inputenc}
\usepackage[dvipsnames]{xcolor}
\usepackage{soul}
\usepackage{fullpage}
\usepackage{tikz}
\usepackage{tikz-cd}

\newcommand{\be}{\begin{equation}}
\newcommand{\ee}{\end{equation}}
\numberwithin{equation}{section}

\newcommand{\Fa}{\mathcal{F}}

\definecolor{vub}{RGB}{0,52,154}
\definecolor{vubo}{RGB}{255,102,0}
\definecolor{redd}{RGB}{255,40,40}
\definecolor{r}{RGB}{228,32,20}
\definecolor{o}{RGB}{238,69,4}
\definecolor{y}{RGB}{253,228,1}
\definecolor{g}{RGB}{108,160,0}
\definecolor{b}{RGB}{0,162,203}
\definecolor{i}{RGB}{120,42,117}

\usepackage[utf8]{inputenc}
\usepackage[framemethod=tikz]{mdframed}
\usetikzlibrary{calc}
\usepackage{chngcntr}

\counterwithin{ex}{section}

\makeatletter
\def\mdf@@mynote{}
\define@key{mdf}{mynote}{\def\mdf@@mynote{#1}}

\mdfdefinestyle{mystate}{
skipabove={1\baselineskip},
linewidth=0.5pt,
innertopmargin=1\baselineskip,
roundcorner=10pt,
backgroundcolor=black!0,
linecolor=black,
singleextra={
  \node[xshift=10pt,draw=black,fill=white,rounded corners,anchor=west] at (P-|O) %
  {\strut{\itshape  }\ifdefempty{\mdf@@mynote}{}{\itshape\bfseries \mdf@@mynote}};
},
firstextra={
  \node[xshift=10pt,draw=black,fill=white,rounded corners,anchor=west] at (P-|O) %
  {\strut{\itshape  }\ifdefempty{\mdf@@mynote}{}{\itshape\bfseries \mdf@@mynote}};
}
}

\mdfdefinestyle{mystateb}{
skipabove={1\baselineskip},
linewidth=1pt,
innertopmargin=1\baselineskip,
roundcorner=10pt,
backgroundcolor=white!5,
linecolor=vub!50,
singleextra={
  \node[xshift=10pt,thick,draw=vub!50,fill=b!0,rounded corners,anchor=west] at (P-|O) %
  {\strut{\itshape  }\ifdefempty{\mdf@@mynote}{}{\bf\mdf@@mynote}};
},
firstextra={
  \node[xshift=10pt,thick,draw=vub!50,fill=b!0,rounded corners,anchor=west] at (P-|O) %
  {\strut{\itshape  }\ifdefempty{\mdf@@mynote}{}{\bf\mdf@@mynote}};
}
}

\mdfdefinestyle{mystater}{
skipabove={1\baselineskip},
linewidth=1pt,
innertopmargin=1\baselineskip,
roundcorner=10pt,
backgroundcolor=white!5,
linecolor=r!50,
singleextra={
  \node[xshift=10pt,thick,draw=r!50,fill=b!0,rounded corners,anchor=west] at (P-|O) %
  {\strut{\itshape  }\ifdefempty{\mdf@@mynote}{}{\bf\mdf@@mynote}};
},
firstextra={
  \node[xshift=10pt,thick,draw=r!50,fill=b!0,rounded corners,anchor=west] at (P-|O) %
  {\strut{\itshape  }\ifdefempty{\mdf@@mynote}{}{\bf\mdf@@mynote}};
}
}

\mdfdefinestyle{mystateg}{
skipabove={1\baselineskip},
linewidth=1pt,
innertopmargin=1\baselineskip,
roundcorner=10pt,
backgroundcolor=white!5,
linecolor=g!50,
singleextra={
  \node[xshift=10pt,thick,draw=g!50,fill=g!0,rounded corners,anchor=west] at (P-|O) %
  {\strut{\itshape  }\ifdefempty{\mdf@@mynote}{}{\bf\mdf@@mynote}};
},
firstextra={
  \node[xshift=10pt,thick,draw=g!50,fill=g!0,rounded corners,anchor=west] at (P-|O) %
  {\strut{\itshape  }\ifdefempty{\mdf@@mynote}{}{\bf\mdf@@mynote}};
}
}

\newmdenv[style=mystate,nobreak=true]{state}
\newmdenv[style=mystater,nobreak=true]{stater}
\newmdenv[style=mystateg,nobreak=true]{stateg}
\newmdenv[style=mystateb,nobreak=true]{stateb}

\makeatother

\makeatother

\newcommand\maps{{\rm Maps}}
\newcommand\redd{/\!\!/}
\newcommand\pd{\partial}
\newtheorem{prop}{Proposition}

\newcommand{\tG}{\widetilde G}
\newcommand{\bbD}{\mathbb D}
\DeclareMathOperator{\ad}{Ad}

\newcommand{\cM}{\mathcal M}

\usepackage{scalerel}
\usepackage{stackengine,wasysym}

\newcommand{\inv}{^{-1}}
\newcommand{\into}{\hookrightarrow}
\newcommand{\onto}{\twoheadrightarrow}

\usepackage{bm}

\definecolor{cambridgeblue}{rgb}{0.64, 0.76, 0.68}
	\definecolor{lapislazuli}{rgb}{0.15, 0.38, 0.61}
\definecolor{awesome}{rgb}{1.0, 0.13, 0.32}
\definecolor{aureolin}{rgb}{0.99, 0.93, 0.0}
\definecolor{almond}{rgb}{0.94, 0.87, 0.8}
\definecolor{antiquewhite}{rgb}{0.98, 0.92, 0.84}

\renewcommand{\AA}{\mathbb{A}}
\newcommand{\hh}{\mathbbm{h}}
\newcommand{\DD}{\mathbb{D}}
\newcommand{\FF}{\mathbb{F}}

\newcommand{\MM}{\mathbb{M}}
\newcommand{\RR}{\mathbb{R}}
\newcommand{\LL}{\mathbb{L}}
\newcommand{\VV}{\mathbb{V}}

\newcommand{\UU}{\mathbb{U}}

\newcommand{\YY}{\mathbb{Y}}
\newcommand{\TT}{\mathbb{T}}

\let\olddigamma\digamma
\renewcommand{\digamma}{\ensuremath \raisebox{1pt}{$\olddigamma$}}

\newcommand{\drf}{Drinfel'd}
\newcommand{\bbg}{\mathbbm{g}}
\newcommand{\tA}{\widetilde A}
\newcommand{\chm}{\check{\mathbbm{m}}_1}

\newtheorem{definition}{Definition}
\newtheorem{cor}{Corollary}

\DeclareMathOperator{\aut}{{\rm Aut}}
\DeclareMathOperator{\inn}{{\rm Inn}}

\newcommand{\Mtot}{\mathbb{M}}

\newcommand{\bbm}{\mathbbm{m}}
\newcommand{\la}{\mathrel{\triangleright}}
\newcommand{\ra}{\mathrel{\triangleleft}}

\newcommand{\bbay}{\mathbbm{a}}
\newcommand{\bbM}{\mathbb{M}}
\newcommand{\bbU}{\mathbb{U}}

\newcommand{\ta}{\tilde{a}}
\newcommand{\bMM}{\bm{\mathcal M}}

\begin{document}

\begin{titlepage}

\vfill

\begin{center}
	\baselineskip=16pt  
	
{\Large \bf  \it A QP perspective on topology change in Poisson-Lie T-duality} 

	\vskip 1cm
	{\large \bf  Alex S.~Arvanitakis$^{a,}$\footnote{\tt alex.s.arvanitakis@vub.be},   Chris D. A. Blair$^{a,}$\footnote{\tt christopher.blair@vub.be}, Daniel C. Thompson$^{a,b,}$\footnote{\tt d.c.thompson@swansea.ac.uk}}
	\vskip .6cm
	{\it  
			$^a$ Theoretische Natuurkunde, Vrije Universiteit Brussel, and the International Solvay Institutes, \\ Pleinlaan 2, B-1050 Brussels, Belgium \\ \ \\
			$^b$ Department of Physics, Swansea University, \\ Swansea SA2 8PP, United Kingdom \\ \ \\}
	\vskip 2cm
\end{center}

\begin{abstract}
\noindent
We describe topological T-duality and Poisson-Lie T-duality in terms of QP (differential graded symplectic) manifolds and their canonical transformations.
Duality is mediated by a QP-manifold on doubled non-abelian ``correspondence'' space, from which we can perform mutually dual symplectic reductions, where certain canonical transformations play a vital role. In the presence of spectator coordinates, we show how the introduction of \emph{bibundle} structure on correspondence space realises changes in the global fibration structure under Poisson-Lie duality. 
Our approach can be directly translated to the worldsheet to derive dual string current algebras.
Finally, the canonical transformations appearing in our reduction procedure naturally suggest a Fourier-Mukai integral transformation for Poisson-Lie T-duality.

\end{abstract}

\vfill

\setcounter{footnote}{0}
\end{titlepage}
\tableofcontents 
 
\newpage
 
\vspace{1em}\noindent

\section{Introduction}

\paragraph{Topological T-duality}
The attraction of studying T-duality is that it is one of the simplest examples of the interesting new features that appear when passing from theories of particles to theories of strings. 
Physicists understand T-duality in practice as a set of transformation rules that take string theory in one spacetime, $M$, admitting $d$ (compact) abelian isometries, to string theory in another spacetime, $\widetilde{M}$, also admitting $d$ abelian isometries.
These transformation rules express how the components of the metric and other fields on $\widetilde{M}$ are determined in terms of those on $M$.
This is therefore at least naively a very local perspective.

A perspective which captures some essential global features of T-duality is provided by topological T-duality \cite{Bouwknegt:2003vb,Bouwknegt:2003zg}.
In the simplest version of this approach, the manifold $M$ is the total space of principal circle bundle with base $B$, with connection $\mathcal{A}$ and curvature $\mathcal{F}$, equipped with a H-flux $\mathsf{H}$ which decomposes as $\mathsf{H} = H + \mathcal{A} \wedge \widetilde{\mathcal{F}}$, where $H$ and $\widetilde{\mathcal{F}}$ are basic forms invariant under the circle action.
The T-dual $\widetilde{M}$ has the same structure but with $\mathcal{F}$ exchanged for $\widetilde{\mathcal{F}}$, and $\mathcal A$ exchanged for a new connection $\widetilde{\mathcal A}$ with curvature $\widetilde{\mathcal F}$. For the original manifold $M$, the connection $\mathcal A$ would be part of the metric data in a Kaluza-Klein ansatz, and the exchange corresponds to the familiar T-duality momentum $\leftrightarrow$ winding swap.
This can further be extended from circle fibrations to the case of a torus bundle.

Topological T-duality can be described in terms of a correspondence space $M \times_B \widetilde{M}$ which combines the original and dual circle fibres into a $T^2$ fibre over $B$. 
Starting from the correspondence space, one has a choice of reductions to either the `original' or `dual' manifold, as illustrated in the following figure:
\be
\label{diagram:circlebundles}
\begin{tikzcd}
& M\times_B\widetilde M\ar[dl,"p"]\ar[dr,"\widetilde p"] &  &T^2\text{-bundle}\\
M\ar[dr,"\pi"]& & \widetilde M\ar[dl,"\widetilde \pi"] &\text{$S^1$-bundles}\\
& B  & &\text{base}
\end{tikzcd}
\ee
Topological T-duality has proven to be of mathematical interest in its own right, while the correspondence space can be seen as a precise mathematical realisation of a `doubled' space, an idea often invoked by physicists in other (non-topological) approaches to T-duality. 

Topological T-duality also interacts with the \emph{Courant algebroid} \cite{courant1990dirac,liu1997manin} structures that describe the gauge algebra of the NS-NS sector of type II supergravity. For example it is known that the exact Courant algebroids atop $M$ and $\widetilde M$ can be quotiented to provide isomorphic ``invariant'' Courant algebroids atop the base $B$ \cite{Cavalcanti:2011wu} which can be seen as another manifestation of topological T-duality. We will be heavily exploiting this algebroid perspective albeit in its equivalent QP-manifold guise, which has already proven fruitful in the study of T-duality \cite{Barmaz:2013yua,Heller:2016abk,Bessho:2015tkk,Carow-Watamura:2018iau,Severa:2018pag}.

\paragraph{Generalisations of T-duality}
Here, we will study an extension of the framework of topological T-duality to treat generalised versions of T-duality. 
The simplest generalisation focuses on manifolds $M$ admitting \emph{non-abelian} isometries \cite{delaOssa:1992vci}, but a more general version of T-duality surprisingly applies in  spacetimes without isometries but which admit a group action of a Poisson-Lie group. This is known as \emph{Poisson-Lie T-duality} \cite{Klimcik:1995ux, Klimcik:1995dy}. Poisson-Lie T-duality is known to be an equivalence of sigma models in the phase space formulation \cite{Sfetsos:1997pi,Klimcik:1995ux, Klimcik:1995dy}, and it contains non-abelian and abelian T-duality as particular cases. 
The difficulty here is that questions of topology are far less tractable. As the dual manifolds in these settings do not involve circle bundles, there is no Gysin sequence to exploit as was done in \cite{Bouwknegt:2003vb,Bouwknegt:2003zg} for the abelian case.  The challenge is further increased by the fact that the dual fibres (roughly speaking the spaces on which the duality acts) are not even diffeomorphic. Despite these challenges, there are many structural similarities between abelian and Poisson-Lie duality. 
 
A crucial feature of Poisson-Lie T-duality is that it admits a naturally doubled interpretation.
This is best understood first in the simplest case $B = \{ \textrm{point} \}$.  Here the target space is simply a group manifold $G$ equipped with the action of a second Lie group $\widetilde{G}$, and the duality transform serves to exchange the role of $G$ and $\widetilde{G}$.  Consistency dictates that given a $G$, $\widetilde{G}$ can not be chosen arbitrarily.  Instead, $G$ and $\widetilde{G}$  are required to be subgroups of a larger group manifold $\mathbb{D}$ whose algebra $\frak{d}$ is equipped with an ad-invariant pairing under-which $\frak{g}= \textrm{Alg}(G)$ and $\tilde{\frak{g}} = \textrm{Alg}(\widetilde{G})$ are maximal isotropic sub-algebras. The Lie group $\mathbb{D}$, known as a classical Drinfel'd double, thus has double the dimension of $G$ and has a similar significance as the doubled $T^2$ fibration in the abelian example above.  

\paragraph{In this paper}
We exhibit changes in the global fibration structure under Poisson-Lie T-duality with spectators. To this end, we extend the diamond diagram \eqref{diagram:circlebundles} from abelian to Poisson-Lie T-duality. At the top of the diamond, we consider the principal $\mathbb{D}$-bundle $\mathbb{D} \into \mathbb{M}  \onto   B$. We then study reductions under the left group actions $ \widetilde G\backslash \mathbb{M}\equiv M$  or $  G\backslash \mathbb{M}\equiv \widetilde{M}$ which take us to the middle rung of the diagram. 
In the presence of spectators, i.e. a non-trivial base $B$, we will find that the bundle $\mathbb{D} \into \mathbb{M}  \onto   B$ can not be arbitrary -- instead, to enable the reduction to $M$ or $\widetilde{M}$ to take place, at least within the approach we are taking here, further conditions need to be imposed.  We show a  sufficient condition for reduction is that $\mathbb{M}$ be a \emph{bibundle} (a right bundle equipped with a right-equivariant structure map) \cite{breen2007bitorseurs,Aschieri:2003mw,murray2012existence} that further more obeys a topological factorisation condition. This factorisation condition requires that $\mathbb{M}$ admits a reduction of its structure group to essentially the center $Z(G)\times Z(\widetilde{G})$.  Though evidently quite stringent, examples are furnished by e.g.~$\mathbb{D} = T^\star G$ and $\mathbb{D} = SL(2, \mathbb{C})$.  

\paragraph{Topological T-duality via QP-manifolds}
Our approach here will be to situate the discussion in the context of QP-manifolds. These are graded symplectic manifolds equipped with cohomological vector field (the Q-structure). This formulation is very practical in the present setting, due to the fact that a Courant algebroid can be very compactly realised as a degree 2 QP-manifold \cite{Roytenberg:2002nu}. The diamond diagram of \eqref{diagram:circlebundles} is upgraded to a diagram of QP-manifolds\footnote{The graded vector bundle $T[1] M$ is the reversed-parity tangent bundle, where the fibre coordinates are declared to be anticommuting and carry degree 1. The bundle $T^\star[P]V$ for some graded vector bundle $V$ first forms the cotangent bundle $T^\star V$ and then shifts degrees on fibre coordinates of $T^\star V$ by $+P$. The notation is standard in the context of supergeometry and AKSZ constructions.}
\be
\label{diagram:qpdiamond}
\begin{tikzcd}
& \bMM\equiv T^\star[2]T[1]\mathbb{M},\,\ar[dl]\ar[dr] & \\ 
 \mathcal M= T^\star[2]T[1]M \ar[dr]& & \widetilde{\mathcal M}= { T^\star[2]T[1]\widetilde M} \ar[dl] \\
& \mathcal M \redd G\cong \widetilde{\mathcal M}/\!\!/\widetilde G  &  
\end{tikzcd}
\ee  
in which the arrows correspond to symplectic reductions that preserve a $Q$-structure. 
The point of view that duality can be studied in terms of reductions of $QP$-manifolds was articulated in the paper \cite{severa2001some} and in the final line of a proceeding, both by \v{S}evera\cite{Severa:1999ay}, and in \cite{Severa:2018pag,vsevera2015poisson} a Courant algebroid perspective 
was developed. Here we will flesh out the idea that Courant algebroid reduction in the graded setup is just a special case of symplectic reduction\footnote{We thank Fridrich Valach for private communication on this point.}.

Our ultimate motivation for this QP-manifold picture is however the possibility of a generalisation to M-theory and U-duality. In that context the algebroid structure of target space is no longer that of a Courant algebroid, but a QP-manifold picture is known \cite{Arvanitakis:2018cyo,Arvanitakis:2019cxy} and symplectic reduction works in the same way.

\paragraph{Canonical transformations and a Fourier-Mukai transformation} 
An important feature of our approach is that the natural Q-structure appearing at the top of the diamond is {\em not} immediately compatible with the symplectic reduction.  Instead  one must first perform a symplectomorphism  of $\bMM$ (i.e.~a canonical transformation which preserves the $P$-structure but which transforms the $Q$-structure).   Roughly speaking the $Q$-structure can be associated with a three-form on $\mathbb{M}$ and the canonical transformation serves to amend this by the exterior derivative of a certain two-form, $\Phi$ ($\widetilde{\Phi}$), to produce a $\tilde{G}$-basic ($G$-basic) three-form that defines a $Q$-structure on $  \mathcal M= \bMM \redd T[1]\tG$ ($\widetilde{\mathcal M}=\bMM\redd T[1]G$).

In the abelian case, we will see that the difference of these canonical transformation $\varpi = \Phi - \widetilde{\Phi}$ plays a vital role in topological T-duality; it provides the kernel for a Fourier-Mukai integral transform \cite{Hori:1999me} that acts on the Ramond-Ramond (RR) sector of the superstring and eventually lifts to the K-theory perspective of T-duality.  In the Poisson-Lie situation we show the same $\varpi$ has an elegant interpretation, in the absence of spectators,  as that of the Semenov-Tian-Shansky symplectic form on $\mathbb{D}$ \cite{Klimcik:1995ux, Klimcik:1995dy}\footnote{An additional role played by $\varpi$ is as the `fundamental 2-form' appearing as part of a para-Hermitian structure in the Born geometry of correspondence space\cite{Svoboda:2020msh}, which has also been studied in the Poisson-Lie setting in \cite{Marotta:2018myj,Marotta:2019eqc}.}. \emph{With} spectators, our techniques give a natural candidate for the Fourier kernel that determines the duality transformation of the RR forms, and we conjecture that it should play a similarly vital role in future developments on the topological nature of PL duality in the context of D-branes. 

\paragraph{From QP-manifolds to the worldsheet}
A final benefit of the QP-perspective that we employ is that it can be directly linked to the associated worldsheet theory. This is a generalisation of the link between  Courant algebroids and worldsheet current algebras established in the seminal work of  Alekseev-Strobl \cite{Alekseev:2004np}.   Here this is achieved through a transgression into the space of maps into ${\cal M}$ from a supermanifold associated to the worldsheet, as in the AKSZ construction \cite{Alexandrov:1995kv}. Recent work by one of us \cite{Arvanitakis:2021wkt} has shown that  a zero-locus reduction procedure can be employed to define a canonical algebra of currents on the world-volume of any brane starting with a $Q$-structure on any QP-manifold ${\cal M}$.  Specialised to the present situation we show how this recovers the current algebra on $L\mathbb{D}$, the loop-group of $\mathbb{D}$, that defines the PL sigma models of Klim\v{c}\'ik and \v{S}evera\cite{Klimcik:1995ux}, and also the PL dual pair of current algebras on $L G$, $L\tG$.  

\paragraph{Outline of this paper}
We structure the paper as follows. 
Section \ref{sec:alsomaster} introduces a QP-manifold atop a doubled or correspondence space for Poisson-Lie T-duality. We describe a general framework for PL duality via symplectic reductions of this QP-manifold, and we show briefly how these reductions give rise to a candidate Fourier-Mukai kernel for the transformation of Ramond-Ramond fluxes. In section \ref{sec:topT} we discuss topological T-duality of principal torus bundles from the QP perspective, and give an analogous discussion for (topological) PL duality in the familiar case where there are no spectators. In section \ref{sec:dualityworldsheet} we outline how our target space considerations produce equivalences of 2-dimensional sigma models. Finally, in \ref{sec:bibundles} we study PL duality including spectators. After some locally-valid motivations, we introduce a class of bibundles with \drf{} double fibres and characterise their existence topologically. We then exhibit certain examples (in section \ref{subsec:examples}) and discuss how the general framework of section \ref{sec:alsomaster} applies to them and furnishes PL duality with spectators. We end with some general remarks and speculations.

\section{A ``master'' QP-manifold on non-abelian doubled space}
\label{sec:alsomaster}
\subsection{The QP non-abelian set-up}
\label{sec:master}
At the top of the diamond diagram, we consider the  principal   $\mathbb{D}$-bundle $ 
\mathbb{D} \into \mathbb{M}  \onto   B$  endowed with a connection one-form
\be
\bbalpha =  - d \mathbbm{ g} \mathbbm{ g}^{-1}    +  \mathbbm{ g}   \mathbb{A}(x)  \mathbbm{ g}^{-1} \, .
\ee 
To obtain a QP perspective we move to the graded supermanifold  $ 
\bMM\equiv T^\star[2]T[1] \Mtot$ which is equipped with a natural Poisson structure in the following way. We can choose local coordinates,  $x^\mu$ on the base and $\mathbb{Y}^I$ on the fibre  both of degree zero,    $\psi^\mu$, $\bbtheta^A$  degree 1  coordinates on the corresponding shifted tangent spaces, and finally $\chi_\mu, \bbxi_A$ (degree 1) and $p_\mu, \mathbbm{p}_A$ (degree 2) coordinates on the shifted cotangent bundle to $T[1] \Mtot$.     With these the symplectic structure on $\bMM$ is given by
\be
\omega = d x^\mu  dp_\mu  + d(\mathbbm{p}_A \LL^A) - d\bbtheta^A d  \bbxi_A  - d \psi^\mu  d \chi_\mu \, .
\label{symp}
\ee
This gives rise to the non-zero Poisson brackets
\be
\label{eq:LIPBs}
\{\mathbb{Y}^I, \mathbbm{p}_A\}= \mathbb{V}_A{}^I\,,\quad\{\mathbbm{p}_A , \mathbbm{p}_B\} = \mathbbm{F}_{AB}{}^C \mathbbm{p}_C \, , \quad \{\bbtheta^A, \bbxi_B \} = \delta^A_B \,, \quad
  \{\chi_\mu , \psi^\nu\} = \delta_\mu^\nu = -  \{p_\mu, x^\nu \}   \, .
\ee
In the above, $\mathbb{L}^A$ and $\mathbb{V}_A$ are left-invariant Maurer-Cartan forms and fundamental vector fields on $\DD$. We can recover a conventional momentum variable by $\mathbbm{p}_I=\LL^A{}_I \mathbbm{p}_A$; this has $\{\mathbb{Y}^I,\mathbbm{p}_J\}=\delta^I_J$.

We can endow $\bMM$  with a standard Q-structure $Q_0 = \{ \Theta_0, \bullet \}$ defined by a (degree 3) Hamiltonian 
\be 
\label{Qzero}
\Theta_0 = -p_\mu \psi^\mu- \mathbbm{p}_A \bbtheta^A + \frac{1}{2}\mathbb{F}_{BC}{}^A \bbtheta^B \bbtheta^C \bbxi_A \, ,
\ee
corresponding to the exterior derivative on $\mathbb M$ which in this trivialisation becomes a sum of the exterior derivative on the base and that on the group manifold fibre. The Jacobi identity ensures that $\{ \Theta_0 , \Theta_0 \} = 0$ and hence that $Q_0$ is a cohomological vector field.    From this perspective the  degree 1 coordinates $\bbxi_A$ generate a global right action on  $T^\star[2]T[1]\mathbb M$ through $\delta_A \bullet = \{ \{\Theta_0, \bbxi_A\} , \bullet\} $ under which the  $\bbtheta^A$  transform adjointly $\delta_A \bbtheta^B = \mathbb{F}_{A C}{}^B \bbtheta^C$.  This thus provides an interpretation of $\bbtheta^A$ as the left-invariant one-forms $\mathbb{L}^A$. In contrast $\hat{\bbxi}  =  \mathbbm{g}^{-1} \bbxi_B \mathbb{T}^B  \mathbbm{g}$ correspond to a basis of vector fields generating left actions, and on $T^\star[2]T[1]\mathbb M$ this is a hamiltonian action, with hamiltonian $Q_0\hat{\bbxi}_A = -\hat{\mathbbm{p}}_A \equiv - \mathbb{U}_{A}{}^I \mathbbm{p}_I$ furnishing the algebra $\{\hat{\mathbbm{p}}_A , \hat{\mathbbm{p}}_B\} =- \mathbb{F}_{AB}{}^C \hat{\mathbbm{p}}_C$ and which leave the $\bbtheta^A$ invariant as $\{\hat{\mathbbm{p}}_A, \bbtheta^B\}=0$.  We emphasise here $\hat{\mathbbm{p}}_A$ is the ``hamiltonian''/``cotangent'' lift of the infinitesimal left $\mathbb{D}$ action, which is well-defined, from which we infer $\hat \bbxi_A$ correspond to a global basis of sections of the vertical bundle.    
 
We then augment the Q-structure with an H-flux $\mathsf{H}_{\Mtot} \in \Lambda^3 T^* \Mtot$ given by
\be
\mathsf{H}_{\Mtot} \equiv H + \tfrac{1}{2} \omega_{CS}(\bbalpha)  \, ,
\ee 
in which $H\in  \Lambda^3 T^*B$ is basic and $\omega_{CS}(\bbalpha )  = \langle \bbalpha  \, \overset{\wedge}{,}\, d \bbalpha  + \frac{2}{3} \bbalpha\wedge \bbalpha \rangle$   is the Chern-Simons three-form.  We then define the Hamiltonian  
\be
\label{Thetadoubledtotal}
\Theta = \Theta_0  + \mathsf{H}_{\Mtot} \, , 
\ee
which is evaluated explicitly using the decomposition of eq. \eqref{eq:CSlaw} (with the replacement $\mathbb{L}^A \rightarrow \bbtheta^A $ and $\mathbb{A}^A \rightarrow \mathbb{A}^A_\mu \psi^\mu$) as 
\be
\label{eq:Thetadoubledtotal:explicit}
\begin{split}
\Theta&= -p_\mu \psi^\mu- \mathbbm{p}_A \bbtheta^A + \frac{1}{2}\mathbb{F}_{BC}{}^A \bbtheta^B \bbtheta^C \bbxi_A  + \frac{1}{6}H_{\mu \nu \rho} \psi^\mu\psi^\nu\psi^\rho\\
&\qquad +
\tfrac{1}{2}(\mathbb A_A+\bbtheta_A) d\mathbb A^A +  \tfrac{1}{6}  \mathbb A^A \mathbb A^B \mathbb A^C \mathbb{F}_{ABC}-\tfrac{1}{12} \mathbb{F}_{ABC}\bbtheta^A\bbtheta^B\bbtheta^C -\tfrac{1}{4} \mathbb{F}_{ABC}\bbtheta^A\bbtheta^B \mathbb A^C \,.
\end{split} 
\ee
A short calculation  shows that $\{\Theta , \Theta\}=0$ when $d \mathsf{H}_{\Mtot}=0$, or
\be
d H =-\frac{1}{2} d \, \omega_{CS}(\mathbb{A}) =-\frac{1}{2}  \langle \mathbb{F}\,  \overset{\wedge}{,} \, \mathbb{F} \rangle= -\frac{1}{2}\mathbb F^A\wedge\mathbb \FF_A \, .
\ee
Everything we have described is so far completely general as long as we have an adjoint invariant inner product $\langle \,,\rangle$.
Henceforth we specialise to the Drinfel'd double case relevant to Poisson-Lie T-duality.
We therefore assume that $\mathbb{D}$ is even-dimensional and its Lie algebra $\mathfrak{d}$ admits a splitting into two sub-algebras $\mathfrak{g}$ and $\mathfrak{\tilde g}$ which are maximally isotropic with respect to a split signature inner product.
We denote this inner product by $\eta$, such that $\eta_{AB} \equiv \eta(T_A, T_B) \equiv \langle T_A, T_B\rangle$ has components $\eta_{ab} = 0= \eta^{ab}$, $\eta_a{}^b = \eta^b{}_a = \delta_a^b$, with generators $T_a$ for $\mathfrak{g}$ and $\widetilde T^a$ for $\mathfrak{\tilde g}$. 
We also make the technical assumption that this is a ``perfect'' double as defined in appendix \ref{paramdd}, which essentially means that one can globally factorise group elements $\mathbbm{g} \in \mathbb{D}$ as $\mathbbm{g} = \tilde g g$ with $g \in G$ and $\tilde g \in \widetilde G$.

\subsection{Conditions for symplectic reduction}
\label{condSympl}

To proceed down to the middle level of the diamond, we now wish to implement a symplectic reduction of the QP-manifold $(\bMM , Q)$ by a $\widetilde{G}$ action. 
We will see however that this can not be done immediately, instead we are required to first perform a canonical transformation described by a two-form $\Phi$ which we shall characterise in general.

If $\widetilde G$ is a subgroup of $\mathbb{D}$, we can consider the quotient under the left group action $ \widetilde G\backslash \mathbb{M}\equiv M$  
which will be the base of a principal $\widetilde G$-bundle $\widetilde G\hookrightarrow\mathbb{M}\onto \widetilde G\backslash\mathbb{M}$ if $\widetilde G\into \mathbb{D}\onto \widetilde G\backslash \mathbb{D}$ is another principal bundle \cite[Proposition 2.1]{rcohenbundlebookunpublished}, which we assume. The left action has a cotangent lift to an action on $\bMM\equiv  T^\star[2]T[1]\mathbb{M}$  infinitesimally generated by a subset of the right-invariant variables $\hat \bbxi_A ,   \hat{\mathbbm{p}}_A$ that we will denote $\hat{\bbxi}^a, \hat{\mathbbm{p}}^a$.
Since
\be
\label{hatconstraintalgebra}
\{ \hat{\mathbbm{p}}_A,\hat{\mathbbm{p}}_B \} =-\mathbb{F}_{AB}{}{}^C\hat{\mathbbm{p}}_C\,,\quad \{\hat \bbxi_A,\hat \bbxi_B\}=0\,,\quad \{  \hat{\mathbbm{p}}_A , \hat \bbxi_B\}=-\mathbb{F}_{AB}{}{}^C \hat \bbxi_C\,,
\ee
the constraints $\hat{\bbxi}^a, \hat{\mathbbm{p}}^a$ form a closed algebra when the generators $\widetilde T^a \in\widetilde{\mathfrak{g}}$ form a Lie subalgebra in $\mathfrak{d}$ (which is the case).
The Q-structure $Q$ --- or more accurately its hamiltonian --- descends\footnote{ See \cite{sniatycki1983reduction} for the classical (ungraded) case, where it is shown that the ring of functions of the reduced phase space seen as a Poisson algebra is isomorphic to $N(\mathcal I)/\mathcal I$ where $ f\in N(\mathcal I)$ iff $\{f,\mathcal I\}\in\mathcal I$. This definition is sensible in the graded case; for that see e.g.~\cite[Proposition 10.1]{cattaneo2013supergeometric}. } to the symplectic reduction when $Q\hat{\bbxi}^a= Q\hat{\mathbbm{p}}^a=0\mod \hat{\bbxi}^a\,, \hat{\mathbbm{p}}^a$.     However,  for $Q=\{\Theta,\bullet\}$ of \eqref{Thetadoubledtotal} we calculate 
\be
\label{Qhatxi}
Q\hat \bbxi^a =- \hat{\mathbbm{p}}^a + \tfrac{1}{2} d\bbalpha^a \,.
\ee
Note that here we make explicit use of the Drinfel'd double inner product used in the Chern-Simons term.
The presence of the $d\bbalpha^a$ term in \eqref{Qhatxi} clearly obstructs the reduction of the Q-structure.

To fix this, we need to consider an equivalent Q-structure $\bar{Q} $ related to $Q$ by the canonical transformation $\bar{Q} \equiv e^\Phi Q e^{-\Phi}$.
where $\Phi$ is a function of degree 2 on $T[1]\mathbb{M}$ (equivalently, a 2-form on $\mathbb{M}$) chosen to ensure that $\bar{Q} \hat{\bbxi}^a= \bar{Q} \hat{\mathbbm{p}}^a=0\mod \hat{\bbxi}^a\,, \hat{\mathbbm{p}}^a$.   Since $T[1]\mathbb{M}$ is a Lagrangian submanifold in $T^\star[2]T[1]\mathbb{M}$, we have $\bar{Q} \equiv\{\bar{\Theta} ,\bullet\}$ with
\be
\bar{\Theta}\equiv e^\Phi \Theta= \Theta +\{\Phi,\Theta\}+0=\Theta-Q\Phi=\Theta-d\Phi\,.
\ee
In the last equality we identified the action of $Q$ on $T[1]\mathbb{M}$ with the exterior derivative on $\mathbb{M}$. 
Using formula \eqref{Qhatxi} we arrive at the equivalent condition explicitly involving $\Phi$:
\begin{align}
\bar{Q} \hat \bbxi^a&=-\hat{\mathbbm{p}}^a + \tfrac{1}{2} d\mathbb{\alpha}^a-\{Q\Phi,\hat \bbxi^a\} \nonumber \\
&=-\hat{\mathbbm{p}}^a+\tfrac{1}{2} d\bbalpha^a -\iota_{\mathbb{U}^a} d\Phi \overset{!}{=} -\hat{\mathbbm{p}}^a \,,
\end{align}
where we identified the contraction $\iota_{\mathbb{U}^a}$ against the right-invariant vector fields $\mathbb{U}^a$ on $\mathbb{M}$ that correspond to $\widetilde{\mathfrak{g}}$ with the Poisson bracket $ \{\hat \bbxi^a,\bullet\}$.  
Therefore the condition  for the reduction of the Q-structure $\bar{Q}$ is simply given by the existence of $\Phi$ such that
\be
\label{QreducibPhicond}
\iota_{\mathbb{U}^a} d\Phi=\tfrac{1}{2} d \bbalpha^a\,.
\ee
Notice that performing the canonical transformation amounts to shifting the closed 3-form in the  Hamiltonian such that $\bar{\Theta}=\Theta_0+\bar{\mathsf{H}}_{\mathbb M}$ with 
\be
\label{Hbar}
\bar{\mathsf{H}}_{\mathbb M}=\mathsf{H}_\mathbb{M} - d\Phi\, . 
\ee
The reduction condition \eqref{QreducibPhicond} has a natural implication for this three-form namely that it be   $\widetilde G$-basic (i.e. $\widetilde{G}$ invariant and horizontal).  Horizontality follows from  
\be
 \iota_{\mathbb{U}^a} \bar{\mathsf{H}}_{\mathbb M} = \frac{1}{2}  \iota_{\mathbb{U}^a}  \omega_{CS}(\bbalpha) -  \iota_{\mathbb{U}^a} d \Phi = \frac{1}{2}  \left( \iota_{\mathbb{U}^a}  \omega_{CS}(\bbalpha) - d \bbalpha^a  \right) = 0  \, ,
\ee
which together with $d\bar{ \mathsf{H}}_{\mathbb M}  =0 $ gives invariance   $\mathcal L_{\mathbb{U}^a}\bar{\mathsf{H}}_{\mathbb M} =0$.    Hence $\bar{\mathsf{H}}_{\mathbb M}$  can be thought of as the pull back of a conventional three-form of  $ \Lambda^3 T^\star(M)$ that we shall denote simply as $\mathsf{H}$. 

It remains to produce an explicit $\Phi$ which obeys \eqref{QreducibPhicond}.
A sufficient condition that ensures this is for $\Phi$ to be $\widetilde{G}$ invariant and obey
\be
\label{eq:necconditions}
{\cal L}_{\mathbb{U}^a} \Phi = 0 \, , \quad \iota_{\mathbb{U}^a} \Phi = - \tfrac{1}{2}   \bbalpha^a \, . 
\ee
This motivates an ansatz of the form
\be
\Phi = \frac{1}{4} \Omega_{CD} \bbalpha^C \bbalpha^D \, . 
\label{OmegaAnsatz1}
\ee
From the second condition of \eqref{eq:necconditions}, we immediately have that $\Omega^{ab}= 0$, $\Omega^a{}_b = - \Omega_a{}^b =  - \delta^a_b$, such that
\be
\Phi = \frac{1}{4} \Omega_{ab} \bbalpha^a \bbalpha^b  + \frac{1}{2} \bbalpha^a \bbalpha_a \, . 
\label{OmegaAnsatz2}
\ee
Since  ${\cal L}_{\mathbb{U}^a} \bbalpha^B = - \mathbb{F}^a{}_C{}^B \bbalpha^C$ the first condition of \eqref{eq:necconditions} requires that, using the form of the Drinfel'd double structure constants (see appendix \ref{notcon}) 
\be\label{eq:Omegadiffeq}
\iota_{\mathbb{U}^a } d \Omega_{bc} +2 f_{bc}{}^a -2 \Omega_{d [b}\widetilde{f}^{ad}{}_{c]} = 0 \, . 
\ee
At this point, the form of $\Omega_{ab}$ and $\Phi$ may be suggestive of familiar structures in generalised T-duality.
In the subsequent sections of this paper, we will make this precise, and obtain the conditions defining $\Omega_{ab}$, by considering in succession the cases where $\mathbb{D}$ is abelian (corresponding to abelian T-duality), where $\mathbb{D}$ is non-abelian and we assume we are dealing with a trivial bundle over a point (corresponding to Poisson-Lie T-duality without spectators), and finally the full non-abelian case with a non-trivial \emph{bibundle} structure (corresponding to Poisson-Lie T-duality with spectators).

\section{Topological generalised T-duality via QP-manifolds}
\label{sec:topT}

\subsection{Topological T-duality}
\label{topT_QP}

We first consider the case where $\mathbb{D}= T^d \times \widetilde T^d$ is abelian and has dimension $2d$. The original QP Hamiltonian \eqref{eq:Thetadoubledtotal:explicit} in this case takes the simple form
\be
\label{ThetaAbel}
\Theta= -p_\mu \psi^\mu- \mathbbm{p}_A \bbtheta^A  + \frac{1}{6}H_{\mu \nu \rho} \psi^\mu\psi^\nu\psi^\rho + \tfrac{1}{2}(\mathbb A^A+\bbtheta^A) d\mathbb A^B \eta_{AB} \,,
\ee
where we have explicitly reinstated the inner product $\eta_{AB}$.
Here there is an obvious action of $O(d,d;\mathbb{Z})$ T-duality ($\YY^A\mapsto \YY'^A = \mathbb{O}^A{}_B \YY^B$ and so on) which is a canonical transformation of the symplectic structure \eqref{symp}, and in particular is a diffeomorphism of the total space $\Mtot$ of the doubled torus bundle. 
This gives a morphism $\mathbb{O}^\star:C^\infty(\bMM)\to C^\infty(\bMM)$ of graded symplectic manifolds (locally determined as $\mathbb{O}^\star \YY^A=\mathbb{O}^A{}_B \YY^B$ etc.), and an isomorphism of QP-manifolds
\be
\mathbb{O}:(\bMM,Q)\to (\bMM,Q')\,,\qquad \mathbb{O}^\star Q=Q'\mathbb{O}^\star
\ee
where the new Q-structure $Q'$ takes the form \eqref{ThetaAbel} in terms of the transformed variables, with $\mathbb{A}'^A = (\mathbb{O}^{-1})^A{}_B \mathbb{A}^B$.


We now complete the symplectic reduction described in section \ref{condSympl}.
We assume the principal $\mathbb{D}=T^d \times \widetilde T^d$ bundle is a principal $T^d$ bundle with respect to the action of both the torus $T^d$ and the dual torus $\widetilde T^d$ inside $T^{2d}$. 
Our symplectic reduction leaves $\Mtot/\widetilde{T}^d\cong M$ as a principal $T^d$-bundle over the base $B$. 
We identify $\alpha^a =\mathcal{A}^a$ with the connection one-form on $M$, with curvature $\mathcal{F}^a = d \mathcal{A}^a$.
We also denote $\alpha_a = \widetilde{\mathcal{A}}_a$ and let $\widetilde{\Fa}_a = d \widetilde{\mathcal{A}}_a$.\footnote{To be precise $\pi^\star \widetilde{\Fa}_a = d \widetilde{\mathcal{A}}_a$ is only a local statement on  $ T^d \hookrightarrow M \overset{\pi}{\rightarrow} B$, but on the dual bundle  $ \widetilde{T}^d \hookrightarrow \widetilde{M} \overset{\widetilde{\pi}}{\rightarrow} B$ we have  $\widetilde{\pi}^\star \widetilde{\Fa}_a = d \widetilde{\mathcal{A}}_a$ globally.}

We need to solve equation \eqref{eq:Omegadiffeq} for the object $\Omega_{ab}$, which in the absence of structure constants becomes simply
$\iota_{\mathbb{U}^a } d \Omega_{bc}  = 0$, and we can take the trivial solution $\Omega_{ab} = 0$. Thus we have $\Phi = \tfrac{1}{2} \bbalpha^a \bbalpha_a = \tfrac{1}{2} \mathcal{A}^a \widetilde{\mathcal{A}}_a$. 
This leads to the QP Hamiltonian and H-flux
\be
\label{abelian:ThetaonM}
\Theta=-p_a\theta^a-p_\mu\psi^\mu +\mathsf{H}\,, \quad
\mathsf{H} = H + \widetilde{\mathcal{F}}_a \mathcal{A}^a\,.
\ee
Alternatively, we could have carried out the `dual' symplectic reduction leading to $\Mtot/{T}^d\cong \widetilde M$.
This is straightforwardly done directly, or can be viewed as arising from the $O(d,d;\mathbb{Z})$ transformations above by first using the transformation that exchanges $\YY^a \leftrightarrow \widetilde \YY_a$ and then reducing as before.
Either way one finds that now $\widetilde{\mathcal{A}}_a$ is the connection one-form on $\widetilde M$, and
\be
\widetilde \Phi = \tfrac{1}{2}\bbalpha_a  \bbalpha^a  =  \tfrac{1}{2}  \widetilde{\mathcal{A}}_a \mathcal{A}^a\,,\quad
\widetilde\Theta = - \widetilde p^a \widetilde \theta_a - p_\mu \psi^\mu + \widetilde{\mathsf{H}}\,,
\quad
\widetilde{\mathsf{H}} = H + \mathcal{F}^a \widetilde{\mathcal{A}}_a \,.
\ee
The exchange of the roles of $\mathcal{A}^a$ and $\mathcal{A}_a$, and hence of $\mathcal{F}^a$ and $\widetilde{\mathcal{F}}_a$, realises topological T-duality \cite{Bouwknegt:2003vb}.

We can continue the reduction procedure to the bottom of the diamond diagram.
Starting with \eqref{abelian:ThetaonM} and reducing by the action of $p_a$ leads to a new Hamiltonian $\Theta = -p_\mu \psi^\mu + H + \widetilde{\mathcal{F}}_a ( \theta^a + A_\mu{}^a\psi^\mu )$.
This is not manifestly invariant under the exchange of $\mathcal{A}^a$ and $\mathcal{A}_a$.
This can be fixed by instead first performing a canonical transformation generated by $\Phi = \xi_a A_\mu{}^a \psi^\mu$ which replaces \eqref{abelian:ThetaonM} with
\be
\Theta = -p_a \theta^a -p_\mu \psi^\mu + H + \widetilde{\mathcal{F}}_a \theta^a + \mathcal{F}^a \xi_a 
\ee
leading to 
\be
\Theta = -p_\mu \psi^\mu + H + \widetilde{\mathcal{F}}_a \theta^a + \mathcal{F}^a \xi_a 
\label{abelianBottom}
\ee
after quotienting by the action of $p_a$. 

We have now arrived at a transitive Courant algebroid whose associated QP-manifold is $\cM\redd T^d=T^\star[2]T[1]B\times \mathbb R^{2d}[1]$, with Q-structure defined via \eqref{abelianBottom}. This is manifestly self-T-dual under the canonical transformation that swaps $\xi_a\leftrightarrow \theta^a$. This is the QP-manifold version of the equivalence of `invariant' Courant algebroids under T-duality \cite{Cavalcanti:2011wu} which was originally noticed by Barmaz \cite{Barmaz:2013yua}.

\subsection{(Topological) Poisson-Lie T-duality}
\label{topPLTnospecs}
 
We now specialise the set-up of section \ref{sec:master} to the case where there are no spectators $B=\{\text{point}\}$, so the correspondence space is a \drf{} double: $\MM=\DD$. In this case the connection $\bbalpha$ on $\MM$ equals the right-invariant Maurer-Cartan form ($\bbalpha=\RR$) and the hamiltonian $\Theta$ of \eqref{eq:Thetadoubledtotal:explicit} on $T^\star[2]T[1]\DD$ simplifies to
\be
\label{ThetaDD}
\Theta=-\bbtheta^A \mathbbm{p}_A + \tfrac{1}{2} \FF_{AB}{}{}^C \bbtheta^A \bbtheta^B \bbxi_C -\tfrac{1}{12} \FF_{ABC}\bbtheta^A\bbtheta^B\bbtheta^C\,.
\ee
A quotient of the \drf{} double $\DD$ by its maximal isotropic subgroup $\tG$ produces the group manifold $G$ (due to the parameterisation $\mathbbm{g}=\tilde g g$). We will upgrade this to a symplectic reduction of the associated QP-manifolds using the general machinery of subsection \ref{condSympl}. This will give rise to Poisson-Lie T-duality. 

We summarise some essential conventions and notations in appendix \ref{notcon}. In particular associated to the decomposition $\mathbbm{g} = \tilde g g$ we introduce the adjoint actions $a$ of $g$ and $\tilde a$ of $\tilde g$ acting on the double. 

Now, the absence of spectators, the canonical transformation described by $\Phi$ takes the form
\be
\label{Phiansatz:nospecs}
\Phi=\frac{1}{2}(\Omega_{ab}\RR^a \RR^b + \RR^a \RR_a)\,, 
\ee
involving the 0-form $\Omega_{ab}$ antisymmetric in its indices.    To solve the condition \eqref{QreducibPhicond} ensuring the symplectic reduction, we can choose
\be
\Omega_{ab}=- 2\widetilde \Pi_{ab}[\tilde{g}^{-1}] \,,
\ee
for $\tilde \Pi_{ab}$ the Poisson bivector on $\widetilde G$ that satisfies \eqref{eq:dPitildeR} and endows it with Poisson-Lie structure.\footnote{For historical reasons we have conventions in which   $\widetilde{ \Pi}_{ab}[ \tilde{g}] $ is  defined to be the Poisson bivector compatible with right multiplication,  and thus   $\widetilde \Pi_{ab}[\tilde{g}^{-1}]$ is  compatible with left multiplication.} 
 
The effect of the canonical transformation generated by this $\Phi$ is to eliminate the term $\FF_{ABC} \bbtheta^A \bbtheta^B \bbtheta^C$ that corresponds to the canonical 3-form on the group manifold $\DD$. To see this we first calculate the adjoint action of $\tilde g$ on the Maurer-Cartan 1-form $\RR=-d\bbg \bbg\inv$ on  $\DD$. Using the parameterisation $\bbg=\tilde g g$ we have
\be
\label{tildeaRR}
\tilde a\RR=\tilde g\inv\RR\tilde g= r+ \tilde l\,.
\ee
This is used to write $\RR$ inside \eqref{Phiansatz:nospecs} in terms of the Maurer-Cartan forms $r,\tilde l$ of $G,\tG$ respectively and the adjoint action $\tilde a$. Writing $\widetilde \Pi$ in terms of components of the adjoint action of $\tilde{g}$ on $\mathfrak{g}$ results in 
\be
\Phi=\tfrac{1}{2} r^a \wedge \tilde l_a\,.
\ee
On the other hand, by identifying $\FF_{ABC} \bbtheta^A \bbtheta^B \bbtheta^C\equiv \FF_{ABC} \LL^A \wedge\LL^B \wedge \LL^C$ and exploiting the adjoint invariance of $\FF$, we can again use \eqref{tildeaRR} to write
\be
\FF_{ABC} \LL^A\wedge \LL^B \wedge\LL^C=3 f_{ab}{}{}^c r^a\wedge r^b\wedge l_c + 3\tilde f^{ab}{}{}_c \tilde l^a\wedge \tilde l^b \wedge r_c
\ee
from which we calculate $\bar \Theta=\Theta-d\Phi$:
\be
\bar\Theta=-\bbtheta^A \mathbbm{p}_A + \tfrac{1}{2} \FF_{AB}{}{}^C \bbtheta^A \bbtheta^B \bbxi_C\,.
\ee
This calculation shows the canonical 3-form on $\DD$ constructed via the $O(d,d)$ structure $\eta$ is exact (see also e.g.~\cite[section 3]{Reid-Edwards:2010plo}).

The general machinery of subsection \ref{condSympl} realises the symplectic reduction
$T^\star[2]T[1]\DD\redd T[1]\tG=T^\star[2]T[1]G$ where both QP-manifolds have the standard Q-structure \eqref{Qzero} (that arises from the de Rham differential) up to canonical transformations.
To display this explicitly, we exploit \eqref{tildeaRR} again to write the Q-structure and symplectic form on $\DD$ in terms of the ones on each isotropic subgroup $G,\tG$. For the fermionic variables we write $\tilde \theta,\tilde \xi$ for the fermions associated to left-invariant forms and vector fields on $\tG$, and $\hat \theta,\hat \xi$ for the fermions associated to the right-invariant ones on $G$. Then we have
\be
\bbtheta ^A \TT_A= a(\hat\theta^a T_a+\tilde \theta_a \tilde T^a)\,,\qquad \bbxi_A \TT^A= a(\hat \xi_a \tilde T^a + \tilde \xi^a T_a)\,.
\ee
From this we can obtain a symplectomorphism $T^\star[2]T[1]\DD \to T^\star[2]T[1] G\times T^\star[2]T[1]\tG$. To do so, let $\omega_{\DD}$ be the symplectic form
\be
\omega_{\DD}= d(\LL^A \mathbbm{p}_A)- d\bbtheta^A d\bbxi_A
\ee
that produces the Poisson brackets \eqref{eq:LIPBs}. 
Similarly we can introduce analogous symplectic forms $\omega_{\tG}$ for $\tG$ (in left-invariant variables) and $\omega_G$ for $G$ (in right-invariant ones):
\be
\omega_{\tG}=d(\tilde l_a \tilde p^a)-d\tilde\theta_a d\tilde \xi^a\,,\qquad
\omega_{G}= d( r^a p^{\mathsf{R}}_a)-d\hat \theta^a d\hat\xi_a\,.
\label{twoomegas}
\ee
By requiring that $\omega_{\DD}=\omega_G+\omega_{\tG}$ we can then solve for the momenta $\tilde p^a,p^\mathsf{R}_a$ on $\tG, G$ respectively. This determines the momenta $\mathbbm{p}_A$ in terms of an affine transformation of the momenta $\tilde p^a,p^\mathsf{R}_a$.\footnote{We have been consistently using hats to denote left-invariant variables that are dressed by an adjoint action. The left-invariant fermions are transformed by the hat into right-invariant ones. However the momenta transform \emph{affinely} instead of linearly when we demand that the symplectic form $\omega_{\hat G}$ in right-invariant variables pulls back to the symplectic form $\omega_G$ in left-invariant ones, so schematically $p^\mathsf{R}_a \tilde T^a=a(p_a \tilde T^a +\theta\xi)=\hat p + \theta\xi\,.$}
In these new momenta variables we have
\be
\label{transformedBarTheta_nospecs}
\begin{split}
\bar\Theta=  -\tilde p^a \tilde \theta_a +\tfrac{1}{2} \tilde f^{ab}{}{}_c \tilde\theta_a \tilde \theta_b \tilde \xi^c - p^\mathsf{R}_a \hat \theta^a -\tfrac{1}{2} f_{ab}{}{}^c \hat\theta^a\hat\theta^b \hat \xi_c\,, 
\end{split}
\ee
In the same new variables, the hamiltonians $\hat\bbxi^a,\hat{\mathbbm{p}}^a=-\bar Q\hat\bbxi^a$ of section \ref{condSympl} that generate the left $T[1]\tG$ action on $\DD$ read
\be
\hat\bbxi^a= \tilde a^a{}_b \tilde \xi^b\,,\qquad \hat{\mathbbm{p}}^a= \tilde a^a{}_c (\tilde f^{bc}{}{}_d \tilde \theta_b \tilde \xi^d + \bar Q \tilde \xi^c)\,.
\ee
Symplectic reduction effectively sets these to zero in the Q-structure and symplectic form. Using also \eqref{transformedBarTheta_nospecs} we see the reduced symplectic form and Q-structure are the canonical ones for $T^\star[2]T[1]G$, namely the standard cotangent bundle form and the Q-structure arising from the exterior derivative:
\be
\label{qpstructure:reduced:nospecs}
\Theta=  - p^\mathsf{R}_a \hat \theta^a -\tfrac{1}{2} f_{ab}{}{}^c \hat\theta^a\hat\theta^b \hat \xi_c\,,\quad \omega_{G}= d( r^a p^{\mathsf{R}}_a)-d\hat \theta^a d\hat\xi_a\,.
\ee

\paragraph{Dual reduction}   If instead of parametrising  $\mathbbm{g}(\mathbb{Y}) =  \tilde{g}(\tilde{y}) g(y)$, we used $\mathbbm{g}(\mathbb{Y}) =  g'(y')\tilde{g}'(\tilde{y}') $ we could similarly perform a  reduction with respect to the left action of $G$.  This procedure invokes a canonical transformation  generated by
 \be
\widetilde{\Phi}= \tfrac{1}{2}\tilde{r}'_a\wedge  l'^a\,.
\ee
This leads to a (primed) dual version of \eqref{qpstructure:reduced:nospecs}, with the dual structure constants $\tilde f^{ab}{}_c$ appearing in place of the original $f_{ab}{}^c$, realising the symplectic reduction 
\be
\label{qpmfld:reduced:nospecs}
T^\star[2]T[1]\DD\redd T[1]G'=T^\star[2]T[1]\widetilde{G}' \,.
\ee
We will sometimes write $G'$ and $\widetilde G'$ instead of $G$ and $\widetilde G$ to clarify when the associated factorisation is $\bbg=g'\tilde g'$ or the original $\bbg=\tilde g g$.

\paragraph{Further reduction and a target-space avatar of the Klim\v{c}\'ik--\v{S}evera current algebra.} The QP-manifold $T^\star[2]T[1]G$ with the QP-structure \eqref{qpstructure:reduced:nospecs} can be reduced again, this time modulo $G$ rather than $T[1]G$. It is convenient to rewrite \eqref{qpstructure:reduced:nospecs} in left-invariant variables:
\be
\label{qpstructure:reduced:nospecsLI}
\Theta=  - p_a  \theta^a +\tfrac{1}{2} f_{ab}{}{}^c \theta^a\theta^b  \xi_c\,,\quad \omega_G= d( l^a p_a)-d \theta^a d\xi_a\,.
\ee
If we directly reduced by the action of $G$ we would end up with a Hamiltonian at the bottom of the diamond diagram which was not invariant under the exchange of $G$ and $\widetilde{G}$.
Therefore we first perform a canonical transformation generated by $\Phi=\tfrac{1}{2}\Pi^{ab}\xi_a\xi_b$ in order to realise a duality-invariant Q-structure later.
For $\Pi$ the Poisson bivector obeying \eqref{eq:dPi} and \eqref{eq:tricky}, we calculate $\{\Phi, Q\Phi\}=0$ and
\be
\label{qstructure:expPhiThetareduced}
e^\Phi \Theta=-p_a \theta^a +\tfrac{1}{2}f_{bc}{}{}^a \theta^b\theta^c \xi_a +p_a\Pi^{ab}\xi_b +\tfrac{1}{2} \widetilde f^{bc}{}{}_a \theta^a\xi_b\xi_c\,.
\ee
In fact by replacing $\Phi\to \lambda \Phi$ for real $\lambda$ this calculation shows that on any Poisson-Lie group $(G,\Pi)$ there is a pair of anticommuting Q-structures on $T^\star[2]T[1] G$: one given by \eqref{qpstructure:reduced:nospecsLI}, and the other by
\be
\Theta_\Pi\equiv p_a\Pi^{ab}\xi_b +\tfrac{1}{2} \widetilde f^{bc}{}{}_a \theta^a\xi_b\xi_c\,.
\ee
Such Q-structures appeared as an example of a ``bi-QP'' manifold in \cite{Grigoriev:2000zg}.
We group the fermionic variables into $\mathbbm{j}^A \TT_A=\theta^a T_a+\xi_a \tilde T^a$, which obey the following suggestive identities given the Q-structure \eqref{qstructure:expPhiThetareduced}:
\be
\label{suggestive}
\{\mathbbm{j}_A,\mathbbm{j}_B\}=\eta_{AB}\,,\qquad \{\mathbbm{j}_A, Q\mathbbm{j}_B\}=-\FF_{AB}{}{}^C \mathbbm{j}_C\,.
\ee
The $\mathbbm{j}_A$ will later give rise to the string sigma-model currents that exhibit Poisson-Lie T-duality, that were originally constructed by Klim\v{c}\'ik and \v{S}evera\cite{Klimcik:1995ux}.  Replacing $G$ with $\tG$ and repeating the above steps leads to another set of $2d$ fermionic variables $\widetilde{\mathbbm{j}}_A$ that obey the same suggestive identities, as long as the Lie algebras $\mathfrak{g},\tilde{\mathfrak{g}}$ form a Manin triple $\mathfrak{d}=\mathfrak{g} \oplus \tilde{\mathfrak{g}}$ with Lie algebra structure constants $\FF_{AB}{}{}^C$.  

We perform the symplectic reduction of $T^\star[2]T[1]G$ with Q-structure \eqref{qstructure:expPhiThetareduced} modulo the left $G$ action, which has generators $-\hat p_a\equiv a_a{}^b p_b$. Since $Q \hat p_a=0\mod \hat p$, the Q-structure descends to the quotient
\be
T^\star[2]T[1]G\redd G=\mathfrak{g}[1]\oplus \mathfrak{g}^\star[1]\,,
\ee 
whose Q and P-structures are obtained from \eqref{qstructure:expPhiThetareduced} and \eqref{qpstructure:reduced:nospecsLI} by setting $ \hat p_a=0$:
\be
\label{eq:Thetabottom}
\Theta= \tfrac{1}{2} f_{ab}{}{}^c \theta^a\theta^b  \xi_c +\tfrac{1}{2} \widetilde f^{bc}{}{}_a \theta^a\xi_b\xi_c=\tfrac{1}{6} \FF_{ABC} \mathbbm{j}^A \mathbbm{j}^B \mathbbm{j}^C\,,\quad 
\omega_G= -d \theta^a d\xi_a=-\tfrac{1}{2} \eta_{AB} d\mathbbm{j}^A d\mathbbm{j}^B\,.
\ee
For this QP-manifold $Q^2=0$ is equivalent to the property that $\mathfrak{d}=\mathfrak{g} \oplus \tilde{\mathfrak{g}}$ is a Manin triple \cite{Grigoriev:2000zg,Voronov:2001qf}. Evidently, had we started with $\tG$ instead of $G$ we would have arrived at an isomorphic QP-manifold, with the replacements $(\theta,\xi)\to(\widetilde \xi,\widetilde \theta)$.

\paragraph{Poisson-Lie duality, canonical transformations, and a lagrangian correspondence.}
The upshot of this subsection is the following diamond diagram of symplectic reductions:
\be
\label{diamonddiagram_nospectators}
\begin{tikzcd}
& T^\star[2]T[1]\DD,\,\ar[dl]\ar[dr] & \\
 T^\star[2]T[1]G \ar[dr]& & { T^\star[2]T[1]\tG} \ar[dl] \\
& \mathfrak{g}[1]\oplus \mathfrak{g}^\star[1]  &  
\end{tikzcd}
\ee 
The top and bottom nodes are manifestly invariant under the symplectomorphism that swaps $ G\leftrightarrow \tG$ in the factorisation $\mathbbm{g}=\tilde g g$ of a $\DD$ group element, which is Poisson-Lie duality. The middle nodes $\cM=T^\star[2]T[1]G$ and $\widetilde{\mathcal M}=T^\star[2]T[1]\tG$ are in general not symplectomorphic because $G$ and $\widetilde{G}$ will not in general be diffeomorphic. However one can view the diagram \eqref{diamonddiagram_nospectators} as implying a \emph{lagrangian correspondence},  i.e. the existence of a lagrangian immersion ($\mathcal L$, $\iota$) consisting of a (graded) submanifold $\mathcal L$ of $\cM\times \widetilde{\mathcal M}$ equipped with the symplectic form $\omega_{\cM\times \widetilde{\mathcal M}}  = \omega_\cM-\omega_{\widetilde{\mathcal M}}$ such that $\iota^\star \omega_{\cM\times \widetilde{\mathcal M}}  = 0$.  Explicitly, we can choose $\mathcal L\equiv T[1](G\times \tG)$ and  $\iota:\mathcal L \to \cM\times \widetilde{\mathcal M}$   acts as (for $\mathbbm{j}^a, \mathbbm{j}_a$ the $\deg 1$ coordinates of $\mathcal L$) $\iota^\star \theta^a=\iota^\star\tilde \xi^a=\mathbbm{j}^a$, $\iota^\star\xi_a=\iota^\star \tilde \theta_a=\mathbbm{j}_a$, $\iota^\star p_a=\iota^\star \tilde p^a=0$ on non-zero degree coordinates.

\section{Bibundles and Poisson-Lie duality with spectators}
\label{sec:bibundles}

We now reinstate the spectator connection field, with the first goal of carrying out the symplectic reduction from the top to the middle rung of the diamond.  The challenge here is to understand the global construction that allows us to solve the general conditions, c.f.~eq.~\eqref{eq:necconditions}, required of the canonical transformation that enables this reduction.   In the absence of spectators the key to doing so was to  parameterise the element of $\DD$ as $\bbg=\tilde g g$ and to then exploit the adjoint action of $\tilde{g}$ on $\DD$ as in eq.~\eqref{tildeaRR}.  In the case of spectators,   in which $\DD$ is potentially non-trivially fibred over $B$  we can proceed on a local patch in much the same fashion, although even here we shall encounter some surprising features.  However to give such manipulations a coherent global meaning we impose additional structure on $\DD \hookrightarrow \MM \twoheadrightarrow B$, namely that it be a bibundle.  

\subsection{A local perspective} 
\label{subsec:local}
To get an intuition of how to proceed, let us fix a local section $\bbg:U\to \DD$ (for open $U\subset B$) such that 
\be  
\bbalpha =  - d \mathbbm{ g} \mathbbm{ g}^{-1}    +  \mathbbm{ g}   \mathbb{A}  \mathbbm{ g}^{-1}  \,, 
\ee
is the  $\frak{d}$-valued connection invariant under local right-acting gauge transformations and equivariant under the global left action.  With $\mathbb{U}_A$ the vectors generating left action, this connection obeys 
\be
\iota_{\mathbb{U}_A} \bbalpha = \TT_A \, , \quad L_{  \mathbb{U}_A}  \bbalpha =- [\TT_A,\bbalpha] \, . 
\ee 
We now want to employ the factorisation $\mathbbm{g} = \tilde{g} g$ and work out the accompanying decomposition of the connection.
A subtle but important point is that the projection of   $\mathbb{A}$ into $\frak{g}$ does {\em not} define a $\frak{g}$-connection.   Instead, we find that   it is the combination $A^a = \mathbb{A}^a +  \mathbb{A}_b \Pi_g^{ba} $ that transforms like a connection under local right $G$ action, i.e.~if  $g\rightarrow g h$ then  $ A^aT_a \rightarrow h^{-1} dh + h^{-1} A^aT_a h $. As a result, we find that the following quantities:
\be
\label{def:rnabla}
r^\nabla = -dg g^{-1} + g A g^{-1} \, ,\quad \tA_a =a_a{}^b   \mathbb{A}_b\,,
\ee       
are invariant under the local right $G$ action. However, in contrast to the abelian case, one cannot directly interpret  $ r^\nabla $ as a connection in a $G$-bundle.  The first way to see this is that the explicit appearance of  $\Pi_g^{ba} $  means   $A^a $ is {\em not} a pull back of a one-form on the base.  More sharply, $r^\nabla$ fails to meet the equivariance condition required of a connection under the global left G action. 
Instead, under the global $G$ left action generated by the vector fields $u_a$ (dual to the right invariant forms $r^a$) we have that\footnote{To establish this result we use 
$
a[g^{-1} ]_{d}{}^b  a[g^{-1} ]_{e}{}^c L_{u_a} \Pi^{de} = - \tilde{f}^{bc}{}_a
$.
}
\be
L_{u_a} (r^\nabla)^b = - f_{ac}{}^b (r^\nabla)^c -  \tilde{f}^{bc}{}_a \tA_c \,  , \quad L_{u_a}   \tA_b= f_{ab}{}^c  \tA_c \, ,
\ee
such that the appearance of $\tA$ in the first equality is the deviation away from $r^\nabla$ being a connection. 

Regardless of these issues, using the decomposition $\mathbbm{g} = \tilde{g} g$ and the definitions \eqref{def:rnabla} we have
\be\label{eq:alphadecomp}
\bbalpha=  \tilde{g} \left[ \tilde{l} +\tA + r^\nabla \right]   \tilde{g}^{-1} \, . 
\ee
We can next evaluate the Chern-Simons form to give (where contractions involving $\eta_{AB}$ are implicit)
 \be
 \label{eq:reducedHflux1}
\begin{aligned}
\langle\bbalpha , d \bbalpha \rangle  + \frac{2}{3} \langle \bbalpha \bbalpha \bbalpha \rangle &=& 
 2 d\Phi    +2 r^\nabla\wedge  r^\nabla \wedge  \tA   +2 r^\nabla\wedge \left[ d\tA +\tA\wedge \tA \right]  
 \end{aligned}  
\ee
in which we have defined (for  $\bar\bbalpha = \tilde{l} + \tA+ r^\nabla$)
\be
\label{phibibundle}
\Phi = \frac{1}{2}  \bar\bbalpha^a \bar\bbalpha_a = \frac{1}{2} (r^\nabla)\wedge (\tilde{l} + \tA) \,  .
\ee
This $\Phi$ takes the form of the ansatz \eqref{OmegaAnsatz1} when expressed in terms of $\bbalpha$. It is manifestly left $\tG$-invariant (because $\bar\bbalpha$ is) and satisfies the conditions \eqref{eq:necconditions}.
The non-exact piece in  eq.~\eqref{eq:reducedHflux1} is evidently $\tG$-invariant and horizontal.   As a result, after canonical transformation and reduction we obtain an H-flux on $\tG\backslash \bbM=M$: 
\be
\mathsf{H} = H    +    r^\nabla\wedge  r^\nabla \wedge  \tA   + r^\nabla\wedge \left[ d\tA +\tA\wedge \tA \right]\, . 
\ee
This provides a local picture of how spectators are to be included but raises critical questions:
\begin{enumerate}
\item In defining the decomposition of eq.~\eqref{eq:alphadecomp} and in solving the conditions for the canonical transformation $\Phi$,  we have assumed three things; first that the fibre coordinates can be split as $\mathbbm{g} = \tilde{g} g$, second that the adjoint action $\bbay[\bbg]$ exists on a bundle,  and third the adjoint action can factorise to give a well-defined maps $\tilde{a}[\tilde{g}]$ and the bi-vector $\tilde{\Pi}$.  What structure needs to be imposed on $\DD\into \MM \onto B$ for these to exist?
\item What is the global status of $r^\nabla$ and $\tA$, and how should they be better understood? 
\end{enumerate} 
To address both of these points we now introduce the aforementioned bibundle structure on $\DD \hookrightarrow \MM \twoheadrightarrow B$.

\subsection{A class of bibundles with Drinfel'd double fibre}
We begin by presenting the definition(s) of a bibundle (in which $\DD$ need not be a Drinfel'd double) \cite{Aschieri:2003mw,murray2012existence}.
\begin{definition}[``Left-right symmetric'']
A (smooth) principal $\DD$-bibundle is a manifold $\MM$ that is simultaneously the total space of a principal bundle for  the left $\la$ and right $\ra$ actions of a group $\DD$ , such that the actions commute, and have the same orbits, $\DD \backslash \MM =\MM / \DD = B$. 

\end{definition}
In the present context we are interested in obtaining a generalisation of the adjoint map $\bbay=\ad \bbg\inv$ to the case of a non-trivial fibration, hence we employ the equivalent definition:
\begin{definition}[``Right action plus (right) structure map'']
\label{rightbidef}
Let $\DD\into \MM \onto B$ be a principal bundle for a right $\DD$ action $\ra$ along with a map (the structure map)
\be
\bbay : \MM \to {\rm Aut}(\DD)
\ee
which is right-equivariant in the sense
\be
\label{rightstructurerightequi}
\bbay[\bbm \ra \bbg ]=\ad[\bbg\inv]\bbay[\bbm]\,.
\ee
 $\MM$ is also a principal $\DD$-bibundle for the  left action $\la$, which is defined by
\be
\label{rightstructure:def}
\bbg\la \bbm= \bbm \ra ( \bbay[\bbm] \,\bbg)\,,
\ee
and in particular the structure map is equivariant under the opposite group action:
\be
\label{lequiay}
\bbay[\bbg\la \bbm]= \bbay[\bbm]\ad [\bbg\inv]\, . 
\ee
\end{definition}
A third equivalent definition can be given in terms of a bundle with left action $\la$ plus (left) structure map  $\hat \bbay[\bbm]$, related to the previous by  $\hat \bbay[\bbm]=(\bbay[\bbm])\inv$ where  inverse denotes the inverse element in $\aut(\DD)$.      In our discussion we will restrict attention to {\em Type 1} bibundles in the terminology of \cite{murray2012existence}. We give an economic characterisation thereof, alternative to the one of \cite{murray2012existence}\footnote{To be absolutely clear, by \emph{Type 1} bibundles we mean the ($\aut(\DD), \DD$) bibundles of \cite{murray2012existence} whose type map $\MM\to \mathrm{Out}(\DD)$ equals 1 everywhere. This implies the existence of a principal $Z(\DD)$-subbundle explicitly given by equation \eqref{type1point}; we refer to that paper for more details. Going the other way around, the existence of the reduction of the structure group to $Z(\DD)$ implies the existence of a $Z(\DD)$ subbundle $\MM_1$, which can be used to construct a left $\DD$ action on $\MM$ (if we are initially given a right one) such that \eqref{type1point} is satisfied.}: 
\begin{definition}[``Type 1 bibundle'']
\label{def:type1}
A principal right $\DD$-bundle $\DD\into \MM \onto B$ is said to be a type 1 bibundle if it admits a reduction of its structure group onto the centre $Z(\DD)$.
\end{definition}

We now specialise to the case where $\DD$ is a Drinfel'd double.  In addition to allowing for the extra structure provided by a bibundle, we want to allow the factorisation $\bbg = \tilde{g} g$ to be applied.   For this we need to introduce a topological condition:
\begin{definition}
\label{topologicalfactorisationcondition} The principal bibundle $\bbM$ of type 1 and with Drinfeld double fibre $\bbD$ obeys the \emph{topological factorisation condition} if $\bbM$   admits a further reduction of its structure group to $Z(\tG,\bbD)\times Z(G,\bbD) $  where
\be  \label{ZGcommabbD:def} Z(G,\bbD)=\{g\in Z(G)| \iota(g)\in Z(\bbD)\} \, , \quad Z(\tilde{G},\bbD)=\{\tilde{g} \in Z(\tilde{G})| \iota(\tilde{g})\in Z(\bbD) \} \, ,  \ee
provide the  central elements in $G$ (resp.~$\tilde{G}$) which are also central in $\DD$.
\end{definition}

The reason now for this definition is it provides a {\bf factorisation of the structure map} 

\begin{prop}[Factorisation of the structure map]
\label{factorisation:prop}
 Let $\bbM$ be a bibundle of type 1 with fibre a perfect Drinfeld double $\bbD$ and structure map $\bbay$ and let $Z(\mathbb{D})\into\bbM_1\onto B$ be the associated $Z(\bbD)$-bundle. If $\bbM$ obeys the topological factorisation condition (Definition \ref{topologicalfactorisationcondition}), then there exist
\be
a: \bbM \to \inn \bbD\,,\quad \tilde a:\bbM \to \inn \bbD\,,
\ee
that respectively fix the subgroups $G,\tG$ inside $\DD$,
\be
\label{ata:fixation}
a[\bbm](\iota G)\subseteq \iota G \,,\qquad \tilde a[\bbm](\tilde\iota \tG)\subseteq \tilde\iota \tG\,,
\ee
and which yield a factorisation of $\bbay$:
\be
\bbay[\bbm]=a[\bbm]\tilde a[\bbm]\,.
\ee
Moreover $a,\tilde a$ enjoy the following equivariance properties involving the inclusions $\iota,\tilde \iota$ of $G,\tG$ in $\bbD$ and the actions $\la,\ra$ of the latter on $\bbM$:
\be
\label{ata:equi}
a[\bbm \ra\iota g](\bbg)=\ad[\iota g\inv] a[\bbm](\bbg)\,,\quad \tilde a[\tilde\iota \tilde g\la \bbm](\bbg)=\tilde a[\bbm]\ad [\tilde\iota \tilde g\inv](\bbg)\,,\quad g\in G\,,\tilde g\in \widetilde G\,, \bbg \in \bbD\,.
\ee
\end{prop}
Note that this proposition depends on the parameterisation $\bbg=\tilde g g$ on the \drf{} double. The same argument goes through with the opposite choice $\bbg=g'\tilde g'$ (the `primed' parameterisation) of \eqref{primedfactorisation} and yields `primed' structure maps $a',\tilde a'$.  

As a consequence of  \eqref{ata:equi},  $a,\tilde a$ restrict to bibundle structure maps for groups $G,\tG$ (with the same total space $\bbM$): 
\begin{cor}
The following expression defines a left $G$ action $\la$ on $\MM$,
\be
\label{the2ndgauging}
g\la_G \bbm\equiv \bbm \ra a[\bbm](\iota g)
\ee
alongside the right $G$ action $\ra$ induced by the restriction of the right $\bbD$ action:
\be
\bbm \ra_G g \equiv \bbm \ra \iota(g)\, . 
\ee
Similarly, the $\tG$ bibundle structure is given by the group actions
\be
\tilde g \la_{\tG} \bbm\equiv \tilde\iota\tilde g \la\bbm\,,\qquad \bbm \ra_{\tG} \tilde g \equiv \tilde a\inv[\bbm](\tilde\iota\tilde g) \la\bbm\,.
\ee
\end{cor}
\noindent The factorised structure maps depend on the factorisation $\bbg = \tilde{g} g$, and for the trivial bibundle we have $a = \ad[g^{-1}]$ and $\tilde{a} = \ad[\tilde{g}^{-1}]$.

Furthermore these bibundle structures are relevant for the reduced spaces: the quotient $\tG\backslash \MM$ inherits the $G$-bibundle structure of $\MM$. This is in fact a type 1 bibundle $G\into \tG\backslash \MM\onto B$ by Definition \ref{def:type1}, as its structure group $Z(G,\DD)$ \eqref{ZGcommabbD:def} is a subgroup of the centre $Z(G)$.

We postpone the proof of the proposition and corollary to section \ref{subsec:proof} and now turn instead to its consequences.

\subsection{Bi-vectors from bibundles} 
The above construction leads to well-defined maps $\tilde{a}$ and  $\widetilde{\Pi}$ that were invoked in the local construction of $r^\nabla, \tA$ and the canonical transformation $\Phi$ (and, of course, equivalents without tildes).  

Since the structure map $\bbay$, and its factorisations $a$ and $\tilde{a}$, take values in $\textrm{Inn}(\DD)\subseteq\textrm{Aut}(\DD)$  we can take a derivative to define maps, that we shall also call, $\bbay$, $a$, $\tilde{a}$ that take values in $\textrm{Aut}(\frak{d})$.  In terms of the generators $\TT_A$ we have for instance $\tilde{a}[\bbm](\TT_A) \equiv  \tilde{a}[\bbm]_A{}^B \TT_B$.  Under the flow generated by the left $\tilde{G}$ action on $\MM$, (where the minus is conventional)
\be
L_{\bbU^a} \bullet = - \frac{d}{d\epsilon} \exp(\epsilon \widetilde{T}^a \la \bullet )  \, , 
\ee
the equivariance of the (factorised) structure maps given by eq.~\eqref{ata:equi} ensures
\be
L_{\bbU^a}  \tilde{a}(\bullet)  = \tilde{a}( [ \widetilde{T}^a , \bullet]) \iff L_{\bbU^a}  \tilde{a}_B{}^C = \FF^a{}_B{}^D \tilde{a}_D{}^C \,.
\ee
Moreover, condition \eqref{ata:fixation} implies the vanishing of the components $\ta^{ab}$ of ${\tilde a}_A{}^B$. We can therefore write
 \be\label{eqn:adecomp}
{\tilde a}_A{}^B=\begin{pmatrix} \delta_a^c & - \widetilde \Pi^\mathsf{L}_{ac}\\ 0 & \delta^c_a \end{pmatrix}  \begin{pmatrix} {\tilde a}_c{}^b & 0 \\ 0 & {\tilde a}^c{}_b  \end{pmatrix}=\begin{pmatrix} {\tilde a}_a{}^b & {\tilde a}_{ab} \\ 0 & {\tilde a}^a{}_b  \end{pmatrix}\,,\quad \widetilde\Pi^\mathsf{L}_{ab}\equiv {\tilde a}_a{}^c {\tilde a}_{bc}\,,
\ee
where $\widetilde\Pi^\mathsf{L}_{ab}$ is antisymmetric and $ 
\tilde a^a{}_b \tilde a_c{}^b=\delta^a_c$, because $\ta[\bbm]$ is an $O(d,d)$ element.  We then have that 
\be
\label{LUtildePi}
  L_{\mathbb{U}^a} \widetilde \Pi^\mathsf{L}_{bc}=-f_{bc}{}{}^a +2 \widetilde \Pi^\mathsf{L}_{d[b} \tilde f^{ad}{}{}_{c]}\,.
\ee 
Up to a factor of $2$ this is precisely identity \eqref{eq:Omegadiffeq} that leads to symplectic reduction modulo $T[1]\tG$. For the case of no spectators we see $\widetilde{\Pi}^\mathsf{L}$ agrees precisely with $-\widetilde \Pi[\tilde g\inv]$, c.f.~\eqref{eq:dPitildeR}. 

\subsection{Connections from projections}  
\label{subsec:connfromproj}

In this section we extract the relevant objects $r^\nabla$ and $\tA$ from the connection $\bbalpha$.   To invoke the split of $\frak{d} = \frak{g} \oplus \tilde{\frak{g} }$ into its maximal isotropic subalgebras, we introduce the commuting projectors $P: \frak{d} \onto \frak{g}$ and $\widetilde{P}: \frak{d} \onto \tilde{\frak{g}} $.   By composing these projections with the factorised structure maps we obtain new projectors, defined point-wise    
 in $\MM$,\be
\label{projectordefs}
\widetilde P_{\langle\tilde a\rangle}\equiv \tilde a\inv \widetilde P \tilde a   \,,\qquad P_{\langle a\rangle}\equiv a P a\inv  \,,
\ee
which in general do not commute.  Similarly to eq.~\eqref{eqn:adecomp} we define a bivector from (the derivative of) the factorised structure map $a:\bbM\to \ad(\frak{d})$:
\be
\Pi^{ab}\equiv a^{ca}a_c{}^b
\ee
which equals the Poisson bivector on $G$ in the absence of spectators in the convention of \eqref{eq:dPi}. We then see that 
\be\label{eq:Pongens}
\widetilde P_{\langle\tilde a\rangle} \widetilde{T}^a = \widetilde{T}^a \ , \quad \widetilde P_{\langle\tilde a\rangle} T_a = + \widetilde{\Pi}^\mathsf{L}_{ab} \widetilde{T}^b \, . 
\ee
Similarly we have 
\be
  P_{\langle a\rangle} \widetilde{T}^a = - \Pi^{ab} T_b   \ , \quad   P_{\langle a\rangle} T_a =  T_a \, . 
\ee 
Using these we can extract the {\em globally well-defined} object 
\be
\label{def:rnabla}
 P  \tilde{a} \bbalpha \equiv r^\nabla \approx r + a^{-1}   P_{\langle  a\rangle} \mathbb{A}
\ee
in which $\approx$ corresponds to the evaluation in a local trivialisation of the kind described in \eqref{checkbbmparameterisation}, in which we can identify (see section \ref{subsec:proof})
\be
\ad[\widetilde g\inv]\approx \ta\,.
\ee
We can now identify the combination
\be
A = P_{\langle\tilde a\rangle} \mathbb{A}= T_a ( \mathbb{A}^a + \mathbb{A}_b \Pi^{ba} )\, . 
\ee
Turning to $\tA$, let us first define the 1-form $\widetilde{\mathcal A}$ on $\bbM$ via  
\be
\label{def:caltildeAdef}
\tilde{a}  \widetilde{ {\cal A} } \equiv \widetilde{P} \tilde{a} \bbalpha  = \tilde{a} \widetilde{P}_{\langle\tilde a\rangle} \bbalpha =  \tilde{a}  \left( \bbalpha_a + \bbalpha^b \widetilde{\Pi}^\mathsf{L}_{ba}  \right) \widetilde{T}^a   \approx \tilde{l} + \tA\, , \quad
\ee 
where again $\approx$ corresponds to the evaluation in a local trivialisation.  We can interpret $\widetilde{ {\cal A} }$ as a global left connection one-form  for the $\widetilde{G}$ action generated by $\tilde{u}^a \equiv \mathbb{U}^a$  i.e.~it obeys the two conditions 
\be
\iota_{\tilde{u}^a} \widetilde{ {\cal A} }   =  \widetilde{T}^a   \, , \quad L_{\tilde{u}^a}\widetilde{ {\cal A} }  = -[   \widetilde{T}^a  ,\widetilde{ {\cal A} }] \, .
\ee 
The algebraic condition on the connection $\widetilde{\cal A}$ follows immediately from eq.~\eqref{eq:Pongens} since $\iota_{\tilde{u}^a} \widetilde{P}_{\langle\tilde a\rangle} \bbalpha  = \widetilde{P}_{\langle\tilde a\rangle} \iota_{\tilde{u}^a}\bbalpha = \widetilde{P}_{\langle\tilde a\rangle} \widetilde{T}^a$. The differential follows easily as for $x\in \frak{d}$, a constant element of the algebra, we have 
\be
\begin{split}
 L_{\tilde{u}^a} \widetilde{P}_{\langle\tilde a\rangle}(x) =&  - \tilde{a}^{-1} ( L_{\tilde{u}^a} \tilde{a}) \widetilde{P}_{\langle\tilde a\rangle}(x) +   \widetilde{P}_{\langle\tilde a\rangle} \tilde{a}^{-1}( L_{\tilde{u}^a} \tilde{a})(x)  \\
 =&   - \tilde{a}^{-1} \tilde{a}( [ \widetilde{T}^a ,  \widetilde{P}_{\langle\tilde a\rangle}(x)] ) +  \widetilde{P}_{\langle\tilde a\rangle} \tilde{a}^{-1}  \tilde{a}( [ \widetilde{T}^a , x])\\
 =& - [ \widetilde{T}^a ,  \widetilde{P}_{\langle\tilde a\rangle}(x)] +   \widetilde{P}_{\langle\tilde a\rangle} ( [ \widetilde{T}^a , x])\, .
 \end{split}
 \ee
Thus 
\be
\begin{split}
 L_{\tilde{u}^a}  \widetilde{ {\cal A} }   =&  ( L_{\tilde{u}^a}  \widetilde{P}_{\langle\tilde a\rangle})\bbalpha  +  \widetilde{P}_{\langle\tilde a\rangle}  L_{\tilde{u}^a}  \bbalpha \\
 =&  - [ \widetilde{T}^a ,  \widetilde{P}_{\langle\tilde a\rangle}(\bbalpha)] +   \widetilde{P}_{\langle\tilde a\rangle} ( [ \widetilde{T}^a , \bbalpha])  -   \widetilde{P}_{\langle\tilde a\rangle} ( [ \widetilde{T}^a , \bbalpha])  =  -[   \widetilde{T}^a  ,\widetilde{ {\cal A} }] \, . 
 \end{split}
\ee
It is important to stress that $\widetilde{ {\cal A} }$ is a connection for a $\tilde{G}$-bundle over $M = \tilde{G}\backslash \MM$ (rather than over the base $B$).  The aforementioned $\tA$ can thus be viewed as the {\em local} gauge potential for this connection.   We are left then with a small puzzle in that the reduced H-flux derived from \eqref{eq:reducedHflux1} ($\eta$ contractions remain implicit)
\be
\label{eq:reducedHflux2}
\mathsf{H}= H +  r^\nabla\wedge r^\nabla \wedge \tA + r^\nabla \wedge [d\tA + \tA\wedge \tA]
\ee
depends on $\tA$ and so appears at first sight to be only a local construction.  However, the combination 
\be
d\tA + \tA\wedge \tA = \ta \widetilde{\cal F}  \, ,
\ee
is globally defined as it is the combination of the curvature 2-form $\widetilde{\cal F}$ on $\MM$ and the globally defined $\ta$.   The other appearance of $\tA$ is in the term $ 
\langle r^\nabla \overset{\wedge}{,} \, r^\nabla \wedge \tA \rangle $  which can be written as  $ 
-\frac{1}{2}(r^\nabla)^a \wedge\langle r^\nabla \overset{\wedge}{,} \,   [T_a, \tA]\rangle  $.  Rather remarkably $[\bullet, \tA]$ {\em is} globally defined even though $\tA$ itself is local. This is because, again assuming a local trivialisation of the form \eqref{checkbbmparameterisation}, the transition functions take values in $Z(\tG)$, and $Z(\mathfrak{g})$ drops out of commutators, so $[\bullet, \tA]$ is gauge-invariant.



\subsection{Proof of factorisation of the structure map (Proposition \ref{factorisation:prop})}
\label{subsec:proof}

The topological factorisation condition of Definition \ref{topologicalfactorisationcondition} is equivalent to the existence of a certain principal bundle of total space $\check{\bbM}_1$, base $B$, and (abelian) structure group $Z(\tG,\bbD)\times Z(G,\bbD)$, that fits into the diagram below:
\be
\begin{tikzcd}
 \check{\bbM}_1 \ar[dr]\ar[r,"\check\iota"] & \MM_1 \ar[d] \ar[r,"\iota_1"] &\MM \ar[dl] \\
  &B&
\end{tikzcd}
\ee
In this diagram $\MM_1$ is the principal bundle with structure group the centre $Z(\DD)$ of $\DD$ whose existence is equivalent to the bibundle $\MM$ being of \emph{type 1} (Definition \ref{def:type1}). This has a concrete realisation as the locus of points $\bbm_1\in \MM$ where left and right actions are equal \cite{murray2012existence}:
\be
\label{type1point}
\bbay[\bbm_1]=1 \iff \bbg \la \bbm_1 = \bbm_1 \ra \bbg \quad \forall \bbg\in \DD\,.
\ee
The map $\iota_1$ is the inclusion of these points into $\MM$. It is a morphism of principal bundles (mapping fibres to fibres and equivariant with respect to the $Z(\DD)\subset \DD$ action). The bundle $\check \MM_1$ is the reduction of $\MM_1$ to the subgroup $Z(\tG,\bbD)\times Z(G,\bbD)\subseteq Z(\DD)$, and $\check\iota$ is the associated morphism of principal bundles. One standard implication of these reductions is that the transition functions of $\MM$ take values in the subgroup $Z(\tG,\bbD)\times Z(G,\bbD)$; see e.g.~\cite[section 2]{nomizu1955reduction}.

For each point $b$ in the base $B$ select an open set $U\subseteq B$ where $\check\MM_1$ is trivial (and thus so is $\MM$). Denote the associated local section of $\check\MM_1$ by $\chm$, so $\chm[b]\in \check \MM_1$ for all points $b\in U$. Then, for all $\bbm \in \MM$ in the fibre above $b\in U$ we can write (omitting the inclusion maps $\iota_1\circ \check\iota$, $\iota,\tilde\iota$)
\be
\label{checkbbmparameterisation}
\bbm= \tilde g g \la\chm[b]
\ee
for unique $\tilde g\in \tG$, $g\in G$ depending on $\bbm,\check\bbm_1[b]$.

\paragraph{Well-definedness of the factorisation.} We can directly calculate the bibundle structure map $\bbay$ in terms of the parameterisation \eqref{checkbbmparameterisation}:
\be
\bbay[\bbm]=\ad[(\tilde gg)\inv]\,,
\ee
and define its factorisation as
\be
a[\bbm]=\ad[\iota g\inv]\,,\qquad \tilde a[\bbm]=\ad[\tilde\iota \tilde g\inv]\,.
\ee
We displayed explicitly the inclusions $\iota:G\to \DD$, $\tilde \iota:\tG\to \DD$ to emphasise that $a[\bbm]$ and $\ta[\bbm]$ are elements of $\aut(\DD)$ rather than $\aut(G),\aut(\tG)$ respectively; in fact they are inner automorphisms.  

To confirm $a,\ta$ are well-defined we choose another local section $\chm':U\to \check \MM_1$, which leads to the new parameterisation (where the primes are unrelated to the `primed' parameterisation of \eqref{primedfactorisation} where $G$ and $\tG$ appear in the other order) 
\be
\bbm=\tilde g' g'\la\check\bbm_1'\,.
\ee
The new section should be related to the old one by a gauge transformation taking values in $ Z(\tG,\bbD)\times Z(G,\bbD) $, so that $\check\bbm_1'[b]=\mathbbm{z}[b]\;\check\bbm_1[b] $ for $\mathbbm{z}[b]\in Z(\tG,\bbD)\times Z(G,\bbD)$, which leads to $\tilde g' \tilde z=\tilde g$, $zg'=g$
once we factorise $\mathbbm{z}=\tilde z z$ in $\tilde z\in\widetilde G$ and $z\in G$. Since $z$ is central in $G$ and $\iota z$ is central in $\bbD$ we find
\be
\ad [g'{}\inv]=\ad [g\inv] \iff a'=a
\ee
and thus $a[\bbm]$ is well-defined. Similarly, $\tilde a[\bbm]$ is also well-defined.

\paragraph{Equivariance.} Now we prove \eqref{ata:equi} for the $a$ map:
\be
a[\bbm \ra\iota g'](\bbg)=\ad[\iota {g'}\inv] a[\bbm](\bbg)\,.
\ee
We again employ the parameterisation \eqref{checkbbmparameterisation} associated to a local section $\chm:U\to \check \MM_1$ of $\check \MM_1$. Then $a[\bbm]=\ad [\iota {g'}\inv]$, and
\be
\bbm\ra\iota g'=( \tilde g g\la \check\bbm_1)\ra g'=\tilde g g\la (\check\bbm_1\ra g')= \tilde g (g g')\la\check\bbm_1\,.
\ee
where we took note of the fact $\chm[b]\in \MM_1$ to employ \eqref{type1point}. Therefore we can read off
\be
a[\bbm\ra\iota g]=\ad [\iota(gg')\inv]= \ad [\iota ({g'}\inv)] \ad [\iota g\inv]= \ad[ \iota ({g'}\inv)] a[\bbm]\,.
\ee
The proof for $\tilde a$ works similarly.

\subsection{Examples}
\label{subsec:examples}

Good examples for our purposes are \emph{non-trivial} bundles $\DD\into \MM\onto B$ that obey the topological factorisation condition of Definition \ref{topologicalfactorisationcondition}. A necessary condition is therefore that \emph{at least one} of the groups $Z(G,\DD),\,Z(\tG,\DD)$ of \eqref{ZGcommabbD:def} is nontrivial, which is to say there must exist elements $z\in Z(G)$ of the centre of $G$ which are also central inside the \drf{} double $\DD$ (resp.~for $\tG$). Then the search for non-trivial such bundles reduces to the classification of the principal bundles with abelian structure group $Z(G,\DD)\times Z(\tG,\DD)$ over the spectator manifold $B$ (these are the $\check \MM_1$ of section \ref{subsec:proof}), which is well-understood. 


\paragraph{Non-abelian dualities.}
A broad class is given by $\DD=T^\star G$, with isotropic subgroups $\tG=\mathfrak g^\star$ and $G$. If $G$ has a non-trivial centre $Z(G)$, we see that $Z(G,\DD)=Z(G)$: the group structure of $\DD=T^\star G$ is the semidirect product $G\ltimes \mathfrak{g}^\star$ where $G$ acts on $\mathfrak{g}^\star=\tilde{\mathfrak{g}}$ via the co-adjoint action, from which it is clear that every $z\in Z(G)$ is also central in $Z(T^\star G)$. On the other hand, the group $Z(\tG,\DD)$ consists of vectors $\widetilde z\in \tilde{\mathfrak{g}}=\mathfrak{g}^\star=\mathbb R^d$ that are fixed points of the co-adjoint action of $G$ on $\mathfrak{g}^\star$. The latter therefore satisfy $\widetilde z( [v,u])=0$ for all $v,u\in\mathfrak{g}$ so they define 1-cocycles for the Lie algebra cohomology $H^1(\mathfrak g)$ \cite[sec.~6.1]{Azcarraga:2011hqa}. Therefore for this class
\be
Z(G,\DD)= Z(G)\,,\qquad Z(\tG,\DD)\subseteq H^1(\mathfrak{g})\,.
\ee
For $G$ compact, connected and semi-simple we fully characterise the (non-)triviality of all bundles involved. With these hypotheses $Z(G)$ is a finite abelian group. Principal bundles with finite group $H$ fibre over connected $B$ are classified by their \emph{monodromy}\footnote{See e.g.~\cite{Freed:1991bn}.}, which is a group homomorphism $\mathsf{m}:\pi_1(B)\to H$. On the other side of the duality, by Whitehead's lemma, $H^1(\mathfrak{g})=0$ so that $Z(\tG,\DD)=1$ and so $\tG\into \widetilde M \onto B$ is always trivial. Therefore non-abelian duality with spectators sends the trivial bibundle $B\times \mathfrak{g}^\star$ to the bibundle $G\into M\onto B$, (non-)trivial depending on the monodromy $\pi_1(B)\to Z(G)$. 

Here we can exhibit the structures that appear in the dual reduced fluxes of the form \eqref{eq:reducedHflux2} (associated to the reduction modulo the left action of $\tG$) and its `primed' analogue (involving the left action of $G$).  Consider firstly the reduction to the left side of the diamond diagram after quotienting by $\widetilde G$. The local quantities appearing are
\be
r^\nabla = - dg g^{-1} + g A g^{-1}  \,,\quad A^a \equiv \mathbb{A}^a \,,\quad \widetilde A_a = a_a{}^b \mathbb{A}_b \,, \quad \Pi^{ab} = 0 \,.
\ee
Hence $r^\nabla$ is in this case a global $G$-connection 1-form, denoted $\mathcal{A}$.  The reduced H-flux is
\be
\mathsf{H} = H + \mathcal{A} \wedge \mathcal{A} \wedge \widetilde A + \mathcal{A} \wedge d \widetilde A\,.
\label{HNATD1}
\ee
The global definitions of the objects appearing here are simply $\mathcal{A} = P \bbalpha$ and $d \widetilde A = \widetilde{\mathcal{F}}$. 

Now consider the reduction to the right side, after quotienting by $G'$. 
The local quantities appearing are
\be
\tilde r'^\nabla = - d \tilde g' \tilde g'^{-1} + \tilde g' \widetilde A' \tilde g'^{-1} \,,\quad
\widetilde A_a' = \mathbb{A}_a + \widetilde \Pi'_{ba} \mathbb{A}^b \,,\quad
A'{}^a = \mathbb{A}^a\,,
\ee
and now
\be
\widetilde{\mathsf{H}}' = H + \tilde r'^\nabla\wedge \tilde r'^\nabla \wedge A' + \tilde r'^\nabla \wedge ( d A' + A' \wedge A')\,.
\label{HNATD2}
\ee
One sees that $A' = A$ and so the final term in the H-flux has exchanged $\widetilde{\mathcal{F}}$ for the field strength $F = dA + A \wedge A$ of the original $G$-bundle $M$.

\paragraph{A Poisson-Lie example.}
Another example is provided by $\DD=\mathrm{SL}(2;\mathbb{C})$, with 
\be
G=\mathrm{SU}(2)\,,\qquad \tG=\mathrm{SB}(2;\mathbb{C})=\Bigg\{\begin{pmatrix} \lambda & z \\ 0 & \lambda\inv \end{pmatrix}\Bigg| \lambda >0\,,z\in \mathbb{C}\Bigg\}\,.
\ee
Clearly $Z(G)=\mathbb{Z}_2$ is central in $\DD=\mathrm{SL}(2;\mathbb{C})$ as well. $Z(\tG)=1$ and $Z(G)$ is finite, so the non-triviality is controlled by possible monodromies $\mathrm{m}:\pi_1(G)\to \mathbb Z_2$ as in the previous example. Duality then relates trivial $\mathrm{SB}(2;\mathbb{C})$ bundles to non-trivial $\mathrm{SU}(2)$ bundles.
 
Here one could again proceed to evaluate the mutually dual H-fluxes $\mathsf{H}$ and $\widetilde{\mathsf{H}}'$ on the Poisson-Lie dual spaces $\mathbb M/\tG$ and $\mathbb M/G$ via formula \eqref{eq:reducedHflux2} although the resulting expressions are not as illuminating, due to the appearance of the somewhat mysterious connection-like objects $r^\nabla, \tilde r'{}^\nabla$ on both sides.

\section{Applications} 

\subsection{Duality on the worldsheet}
\label{sec:dualityworldsheet}

A direct application of the QP-manifold approach we have taken is to immediately obtain a string worldsheet description.
A classic result of Alekseev and Strobl \cite{Alekseev:2004np} shows that the symmetries of a non-linear sigma model are realised through currents specified by a pair of a (pull back of a target space $M$) vector and a one-form, and that these currents form an algebra under the Poisson bracket whose closure invokes the Courant bracket of $TM + T^\star M$. 
The QP perspective was recently found \cite{Arvanitakis:2021wkt} to give a route to this worldsheet result, using a mechanism known as Zero Locus Reduction (ZLR) \cite{Grigoriev:2000zg} alongside the AKSZ construction. As orientation, we first illustrate this in the abelian set-up before proceeding to the case where $\mathbb{D}$ is a Drinfel'd double for which we show the ZLR procedure recovers the Klim\v{c}\'ik--\v{S}evera current algebra.   

Before specialisation to the case at hand we state the key result of this ZLR.   The ingredients of this construction are a QP-manifold $\{{\cal M}, Q,\omega\}$ of degree $p$ and a Q manifold with integral $\{{\cal N}, D,\int_{\cal N}\}$ (the target and source respectively) and the supermanifold of mappings $\textrm{Maps}({\cal N} \rightarrow {\cal M} )$ to which $Q$ and $D$ naturally lift and mutually anticommute. Then $\Delta = D- Q$ is nilpotent.   For a function $f\in C^\infty ({\cal M})$ we denote its pull-back by $\varphi\in  \textrm{Maps}$ as the superfield $\bm{f}= \varphi^\star f$.  We require further that integration by parts holds:  $0=\int_{\cal N} D(\dots)$.  For a   test function $\epsilon \in C^\infty( {\cal N})$ we define the {\em current}, a functional of maps, as  
\be
\label{currentdef}
\langle f | \epsilon \rangle = (-1)^p \int_{{\cal N} } \bm{f} \epsilon \in C^\infty(\textrm{Maps})\,
\ee
We define the {\em zero-locus} or {\em phase space}, ${\cal Z}_{\Delta}$, to be the quotient of $C^\infty(\textrm{Maps})$ by the ideal ${\cal I}_\Delta$  generated by $\Delta$.  As was shown in \cite{Arvanitakis:2021wkt} ${\cal Z}_{\Delta}$ is equipped with a degree zero Poisson bracket  under which the currents, so constructed, form an algebra
\be
\label{universalbracket}
\{\langle f | \epsilon \rangle , \langle g  | \eta \rangle   \}_{{\cal Z}_{\Delta}} = \pm \langle \{f, Qg\} | \epsilon \eta\rangle  \pm  \langle \{f, g\} | \epsilon D\eta \rangle  
\ee
where the signs are determined by the degrees of $f$, $g$ and $\epsilon$ and $p$.  

In the present context, we will specify ${\cal N} = T[1]S^1$ where the $S^1$ is to be thought of as a spatial section of the string worldsheet parameterised by a coordinate $\sigma$,  and $D$ be the de Rham differential $d$ such that superfields of degree $n$  are expanded as $\bm{f} = f^{(n)}(\sigma) + f^{(n-1)}(\sigma) d\sigma$ and the integration $\int_{\cal N}$ simply picks out the top-form and integrates.  

\paragraph{Abelian currents}
We first display this ZLR procedure in the abelian case starting at the bottom of the diamond.
The Hamiltonian was derived in equation \eqref{abelianBottom}, and the corresponding QP-manifold can be seen to be $T^\star[2]T[1]B\times \mathbb R^{2d}[1]$ with the $P$ and $Q$ structure specified by 
\be
\omega=dp_\mu dx^\mu -d\psi^\mu d\chi_\mu - d\xi_ad\theta^a\,,\quad \Theta=-p_\mu\psi^\mu + H +\widetilde{\mathcal F}_a\theta^a + \mathcal F^a \xi_a\,.
\ee
The corresponding target-space algebroid is the ``transitive'' Courant algebroid of $T^d$-invariant sections of $T M\oplus T^\star M$ or equivalently of $\widetilde T^d$-invariant sections of $T\widetilde M\oplus T^\star\widetilde M$. Since this is not an exact Courant algebroid the Alekseev--Strobl result \cite{Alekseev:2004np} does not apply, and the ZLR technology is needed.  
  
The degree 1 coordinates $\xi$ and $\theta$ correspond to sections of the generalised tangent bundle that point along fibre directions.  These give rise to   corresponding currents  e.g.~
\be
\langle \xi_a|\epsilon\rangle=\int_{T[1]S^1}\bm{\xi}_a \epsilon = \oint d\sigma (  \xi_a^{+1} \epsilon^{+1} + \xi_a \epsilon^{0})  \,,\quad \bm{\xi_a}(\sigma,d\sigma)= \xi_a^{+1}(\sigma)+ \xi_a(\sigma)d\sigma\,,
\ee
with $\xi_a^{+1}(\sigma)$ carrying intrinsic degree $+1$ while $\xi_a(\sigma)d\sigma$ is 1-form in $\Gamma( T^*S^1)$ and where $\epsilon\in C^\infty(T[1]S^1)$ is an arbitrary $S^1$-polyform expanded as $\epsilon = \epsilon^{0}(\sigma) + \epsilon^{+1}(\sigma) d\sigma$.   We pick $\epsilon,\eta$ to be 0-forms which picks out the 1-form components in the current (that we actually want to expose). The bracket formula \eqref{universalbracket} then gives {\bf for any background $H,\mathcal{F},\tilde{\mathcal F}$}
\be
\{\langle \xi_a|\epsilon\rangle\,, \langle \theta^b|\eta\rangle\}=-\delta^b_a\int_{T[1]\Sigma} \epsilon d\eta \implies \{\theta^b(\sigma_1)\,, \xi_a(\sigma_2)\}=  + \delta^b_a \frac{\partial}{\partial \sigma_2}\delta(\sigma_1-\sigma_2)\,.
\ee
This motivates considering the currents $\langle \mathbbm{j}^A | \epsilon \rangle$ in which    $ \mathbbm{j}^A    \TT_A=\bm{\theta}^a T_a+\bm{\xi}_a \tilde T^a$ are previously introduced combinations but now viewed as superfields.   The one-form component of $ \mathbbm{j}^A$, which we also call $\mathbbm{j}^A$, then obey {\em worldsheet} Poisson bracket relations 
 \be
\{\mathbbm{j}_A(\sigma_1) ,\mathbbm{j}_B (\sigma_2) \}=\eta_{AB} \delta^\prime(\sigma_1-\sigma_2)  \,.
\ee
The non-zero righthand side is the classical precursor of Schwinger term in a current algebra.  

For {\bf zero background $H=\mathcal F=\tilde{\mathcal F}=0$} the full ZLR and nonvanishing Poisson brackets are
\be
\begin{split}
d\bm{\xi_a}=0\,,\quad d\bm{\theta^a=0}\,,\quad d\bm{x}^\mu=\bm{\psi}^\mu\,,\quad \bm{p}_\mu=-d\bm{\chi}_\mu\,,\\
\{\theta^b(\sigma_1)\,, \xi_a(\sigma_2)\}= + \delta^b_a \frac{\partial}{\partial \sigma_2}\delta(\sigma_1-\sigma_2)\,,\quad \{X^\mu(\sigma_1), P_\nu(\sigma_2)\}=\delta^\mu_\nu \delta(\sigma_1-\sigma_2)\,,
\end{split}
\ee
Solving the ZLR condition for $ \bm{\xi}_a,\bm{\theta}^a $  as 
\be
\bm{\xi}_a=d\tilde{\bm{y}}_a\,,\quad \bm{\theta}^a=d\bm{y}^a
\ee
exposes the familiar  \cite{Tseytlin:1990nb}   non-commutativity between positions/dual positions,
\be
\{y^b(\sigma_1)\,, \tilde{y}_a(\sigma_2)\}= + \delta^b_a\delta(\sigma_1-\sigma_2) + \text{const.}
\ee 
The spectator coordinates $x^\mu$ and their superfields $\bm{x}^\mu$ give rise to an Alekseev--Strobl type current (sub)algebra for zero background, but for nonzero fluxes the fibre and spectator currents will have nonvanishing brackets determined via \eqref{universalbracket}.
 
\paragraph{Non-abelian currents}

We may extend this to the full non-abelian set up, taking for expediency the base $B =\{ \textrm{point}\}$ such that the relevant QP-manifold is $\mathfrak{g}[1]\oplus \mathfrak{g}^\star[1]$ with  Q and P-structures given as in eq.~\eqref{eq:Thetabottom} namely 
\be
\label{eq:ThetabottomRepeated}
\Theta= \tfrac{1}{2} f_{ab}{}{}^c \theta^a\theta^b  \xi_c +\tfrac{1}{2} \widetilde f^{bc}{}{}_a \theta^a\xi_b\xi_c=\tfrac{1}{6} \FF_{ABC} \mathbbm{j}^A \mathbbm{j}^B \mathbbm{j}^C\,,\quad \omega= -d \theta^a d\xi_a=-\tfrac{1}{2} \eta_{AB} d\mathbbm{j}^A d\mathbbm{j}^B\,.
\ee
Then for the currents $\langle \mathbbm{j}^A | \epsilon \rangle$ with  $\epsilon,\eta$ set as  0-forms   we see very rapidly using the suggestive identities \eqref{suggestive} that 
\be
\begin{aligned}
 \{\langle   \mathbbm{j}_A  | \epsilon \rangle , \langle  \mathbbm{j}_B | \eta \rangle   \}_{{\cal Z}_{\Delta}} &=  \langle \{  \mathbbm{j}_A, \{\Theta ,  \mathbbm{j}_B\}\} | \epsilon \eta\rangle  -  \langle \{  \mathbbm{j}_A ,   \mathbbm{j}_B \} | \epsilon d\eta \rangle  \\
&=   -\FF_{ABC}      \langle  \mathbbm{j}^C   | \epsilon \eta\rangle - \eta_{AB} \langle  1 | \epsilon d\eta \rangle
 \end{aligned} 
\ee
Hence,  the one-form component of $ \mathbbm{j}^A$, which we also call $\mathbbm{j}^A$, then obey on the {\em worldsheet} the Klim\v{c}\'ik--\v{S}evera \cite{Klimcik:1995ux} Poisson bracket relations
 \be
\{\mathbbm{j}_A(\sigma_1) ,\mathbbm{j}_B (\sigma_2) \}= -\FF_{AB}{}^C  \mathbbm{j}_C (\sigma_1) \delta(\sigma_1 -\sigma_2)  - \eta_{AB} \frac{\pd} {\pd \sigma_1}\delta(\sigma_1-\sigma_2) \,.
\ee

\paragraph{Gauging worldsheet currents vs.~symplectic reduction}


The procedure of section \ref{subsec:proof} that yields (topological) PL duality via symplectic reductions of QP-manifolds naturally suggests how to implement duality on the worldsheet. We outline briefly how this works.

Consider firstly gauging a current $J_f(\sigma,t)$ corresponding via \eqref{currentdef} to some arbitrary $f\in C^\infty(\mathcal M)$. In the hamiltonian formalism we would simply introduce a lagrange multiplier $\lambda(\sigma,t)$ and add 
\be
\int dt \oint d\sigma\; J_f(\sigma,t) \lambda(\sigma,t)
\ee
to the action. Upon switching to the notation actually used in \eqref{currentdef}, we would rewrite this as 
$\int dt \;\langle f|\lambda\rangle(t)$ where now the test (poly)form $\lambda$ also depends on time $t$, and $\langle f|$ does as well via the superfield $\bm{f}(t)=\varphi(t)^\star f$, for $\varphi:\mathbb R\to \maps$. Therefore the closure of the gauge algebra is controlled by the bracket formula \eqref{universalbracket}. Moreover we see that if we gauge a current $\langle f|$ associated to $f$, we are simulaneously gauging the current $\langle Qf|$ associated to $Qf$: if $\lambda=d\mu$ (for $d$ the worldspace de Rham) then by integration by parts and zero locus reduction $\langle f|\lambda\rangle= (-1)^{f+1}\langle df|\mu\rangle=(-1)^{f+1}\langle Qf|\mu\rangle $. Consistency is thus controlled by the algebra of both $\langle f|$ and $\langle Qf|$ currents.

We can implement duality on the worldsheet by taking the sigma model corresponding to the doubled/correspondence space QP-manifold $\bMM$ and gauging it in two different ways, \`a la Rocek and Verlinde \cite{Rocek:1991ps}. Here it is useful to consider again no spectators $B=\{\text{point}\}$ and recall the diagram of symplectic reductions of QP-manifolds \eqref{diamonddiagram_nospectators}. The PL-dual sigma models have targets the group manifolds $M=G$ and $\widetilde M=\tG$ respectively, and their phase spaces are obtained by the ZLR procedure from the target QP-manifolds $T^\star[2]T[1] G$ and $T^\star[2]T[1]\tG$. Therefore the gauging procedure on the worldsheet must reflect the symplectic reduction on target space, from $\bMM=T^\star[2] T[1]\DD$ to $T^\star[2]T[1] G$ (resp.~$T^\star[2]T[1]\tG$). The symplectic reduction that gives $T^\star[2]T[1] G$ is the reduction modulo $T[1]\widetilde G$ of section \ref{condSympl}, with generators $\hat \bbxi^a$ in degree 1 and $\hat{ \mathbbm{p}}^a\propto \bar Q \hat \bbxi^a$ in degree 2 (where $\bar Q$ is related to the original $Q$ by a canonical transformation). This and the previous paragraph suggest we gauge just the degree 1 currents, i.e.~we introduce $d$ lagrange multiplier 0-forms (on $S^1$) $\lambda_a$ and add
\be
\int dt\: \langle \hat \bbxi^a| \lambda_a\rangle
\ee
to the action. It is easy to see that this resolves into an expression linear in momenta along $\tG$. The gauge algebra indeed closes due to \eqref{hatconstraintalgebra}.

Clearly the same reasoning works \emph{mutatis mutandis} also for the case with spectators, and the case where we want to obtain the model with target $\widetilde M$ instead of $M$.


\subsection{The Fourier-Mukai kernel}
\label{fm}

\paragraph{Fourier-Mukai kernel from canonical transformations} 

For a Drinfel'd double, we can perform reductions of the fibre $\mathbb{D}$ by either $\widetilde{G}$ or $G'$, using canonical transformations $\Phi$ or $\widetilde{\Phi}$ resulting in the H-flux
\be
\bar{\mathsf{H}}_{\mathbb{M}} = \mathsf{H}_{\mathbb{M}} - d \Phi \,,
\label{Hbarhere}
\ee
which is a pullback of a three-form on $\Lambda^3 T^*(M)$ which we can denote ${\mathsf{H}}$, or the dual H-flux
\be
\widetilde{\bar{\mathsf{H}}}_{\mathbb{M}} = \mathsf{H}_{\mathbb{M}} - d \widetilde \Phi \,,
\label{Hbartilde}
\ee
which is a pullback of a three-form on $\Lambda^3 T^*(M)$ which we can denote $\widetilde{\mathsf{H}}$.
Subtracting \eqref{Hbartilde} from \eqref{Hbarhere}, and leaving the pullback implicit, we have
\be
\mathsf{H} - \widetilde{\mathsf{H}}
 = - d(\Phi - \widetilde \Phi ) \,.
\label{diffH}
\ee
We denote the difference of the canonical transformations by $\varpi = \Phi - \widetilde{\Phi}$.
This can be used to obtain an appropriate generalisation of the Fourier-Mukai kernel 
\be\label{eq:FM1} 
{\bf \Phi}= \int_{G'} e^{\mp \varpi}p^\star \,  : \Lambda^\bullet T^\star(M ) \rightarrow  \Lambda^\bullet T^\star(\widetilde{M} )  \, . 
\ee
The idea here is that we first implement a pull back to the doubled space with $p^\star$ and then a push forward $\tilde{p}_\star$ by integrating over the group  $G'$ in the ``dual'' decomposition $ \mathbb{D} = G'\times \widetilde{G}'$. Note that as submanifolds of $\DD$, $G$ and $G'$ are identical (at least in the case of perfect doubles, see formulas \eqref{originalfactorisation}, \eqref{primedfactorisation} and associated discussion in the appendix).  While this integral transform appears naturally,   further development is required e.g. to provide  a specification of   the integral measure taking into account that  $G^{(\prime)}$ need not be  compact. 
 
In the abelian case, this integral transformation is known to describe the transformation under T-duality of the Ramond-Ramond field strengths of Type II string theory.
In that case, the results of section \ref{topT_QP} immediately give 
\be
\label{def:FMkernelAbelian}
\varpi = \Phi -\widetilde{\Phi} = \mathcal{A}^a \widetilde{\mathcal{A}}_a 
\ee
with $\mathsf{H} - \widetilde{\mathsf{H}} = d ( \mathcal{A}^a \mathcal{A}_a )$, and this indeed reproduces the Fourier-Mukai kernel discussed in the context of topological T-duality in \cite{Bouwknegt:2003vb,Bouwknegt:2003zg}.

We therefore suggest the same holds for Poisson-Lie duality. 
The RR fields are ${\bf G} \in  \Lambda^\bullet T^\star(M )$ and obey the H-twisted Bianchi identity, $(d\pm \mathsf{H}\wedge){\bf G} =0$.
The dual field strength $\widetilde{\bf G} \in  \Lambda^\bullet T^\star(\widetilde M)$ can be defined by 
\be
\widetilde{\bf G}  = \int_G e^{\mp \varpi} {\bf G} 
\ee
and thanks to \eqref{diffH}  obey  $(d\pm \widetilde{\mathsf{H}}\wedge)\widetilde{\bf G} =0$.  Therefore the Fourier-Mukai transform \eqref{eq:FM1} defines a morphism of twisted de Rham complexes whenever the integral makes sense for all forms, e.g.~when $G$ is compact (which can be arranged, for instance, in the non-abelian duality example discussed in subsection \ref{subsec:examples}).

For the full Poisson-Lie cases, we now make the expressions for $\varpi$ more explicit and make some further comments.

\paragraph{Proposed Fourier-Mukai kernel in (spectator-free) topological Poisson-Lie duality}

In section \ref{topPLTnospecs} we described how to perform reductions of the fibre $\mathbb{D}$ by either $\widetilde{G}$ or $G$, using canonical transformations given by
\be
\Phi = \frac{1}{2} r^a \wedge \tilde l_a \,\quad
\widetilde{\Phi}= \tfrac{1}{2}\tilde{r}'_a\wedge  l'^a\,.
\ee
The combination 
 \be\label{eq:STS1}
\varpi= \Phi -  \widetilde{\Phi}  = \tfrac{1}{2} r^a\wedge\tilde l_a - \tfrac{1}{2}\tilde{r}'_a\wedge  l'^a \,.
\ee
is closed and non-degenerate and defines  the Semenov-Tian-Shansky symplectic form on the Drinfel'd double. Indeed, this form appeared in the original work of Klim\v{c}\'ik and \v{S}evera and played a distinguished role in the construction of a Lagrangian for a Poisson-Lie invariant ``doubled'' sigma model.

While elegant, the form of eqn \eqref{eq:STS1} is some what harder to apply in practice since it requires knowledge of the diffeomorphism  $y' =y'(y, \tilde{y} ) $ and $ \tilde{y}'=\tilde{y}'(y, \tilde{y} )$.   To address this,  we expand  the one-forms $\tilde{r}'_a ,  l'^a$ in the basis of $r^a$ and $\tilde l_a$  by equating the left invariant form $\mathbb{L}$ written in the two parametrisations  of  $\mathbbm{g} $.     Doing so and invoking identities that follow from equating the adjoint action of $\mathbbm{g}$ on the Drinfel'd double  written in the two parametrisations one obtains   (with wedge product implicit)
   \be\label{eq:STS2}
2 \varpi=  2 r    {\cal O}^{-1}  \tilde{l}  +   \tilde{l} \Pi_{g^{-1}}   {\cal O}^{-1}  \tilde{l}    - r  {\cal O}^{-1}  \widetilde{\Pi}_{\tilde{g}}    r \, , 
\ee
in which  $\Pi_g$ is the PL bivector of $G$, $\widetilde{\Pi}_{\tilde{g}} $ that of $\widetilde{G}$\footnote{In the convention of \eqref{eq:dPi} and \eqref{eq:dPitilde}. We changed notation so the group element that the bivector depends on is now a subscript ($\Pi_g$ instead of $\Pi[g]$) for space reasons. Exceptionally invoke $-\Pi_{g\inv}$ here which is the analogue of $\Pi_g$ but whose derivative satisfies a right-invariant version of \eqref{eq:dPi}, c.f.~$\widetilde \Pi^\mathsf{L}$ of \eqref{eq:dPitildeR}.}  and ${\cal O} \equiv 1- \tilde{\Pi}_{\tilde{g}}  \Pi_{g^{-1}}$.  

For applications as a Fourier-Mukai kernel it is useful to further recast $\varpi$ in terms of the one-forms  $r$ and $\tilde{r}'$ which would enter the definition of the two T-dual theories.   In this case one obtains 
\be\label{eq:STS3}
2 \varpi=  r \widetilde{\Pi}_{\tilde{g}}  r - \tilde{r}' \Pi_{g'}  \tilde{r}' + 2 r  a_g \mathcal{N}^{-t} \tilde a'_{\tilde g'}{}^t  \tilde{r}'   \, ,
\ee
in which now $ a_{g}$ is the adjoint action of $g$ on $\mathfrak{g}$  and $\tilde{a}'_{\tilde{g}'}$ that of $\tilde{g}'$ on $\tilde{\frak{g}}$, and $\mathcal{N} \equiv 1 - \Pi_g \widetilde{\Pi}'{}_{\tilde g'}$.
 
We can specialise this to the case of semi-abelian Drinfel'd doubles, the context for non-abelian T-duality for which a Fourier-Mukai kernel was proposed in  \cite{Gevorgyan:2013xka}. In this case, $\tilde{\frak{g}} = \frak{u}(1)^d$.   With $\tilde{g} = \exp \tilde{y}_a \tilde{T}^a$ such that $\tilde{l}_a = \tilde{r}_a =-  d \tilde{y}_a$ we have 
 \be
 \Pi_{g'} = 0 \ , \quad   \widetilde{\Pi}_{\tilde{g}}  = + f_{ab}{}^c \tilde{y}_c  \ ,  \quad  \tilde{a}_{\tilde{g}} = \textrm{id}  \, 
 \ee
hence $\mathcal{O} = 1$ and from \eqref{eq:STS2} we obtain 
 \be\label{eq:STS3}
2 \varpi=  - 2 r^a d \tilde y_a   - f_{ab}{}^c \tilde{y}_c   r^a  r^b  \, ,
\ee
which agrees (modulo conventions) with the expression in \cite{Gevorgyan:2013xka}.
In this case the relationships with the primed variables are $\widetilde \Pi = a \widetilde \Pi' a^t$, $\tilde l' = a^{-1}_g \tilde l - \tilde \Pi'_{\tilde g'} l$ and $l'=l$,
such that we could alternatively write
\be
2\varpi = r a_g \widetilde \Pi_{\tilde g'} a_g^t r + 2 r a_g \tilde r' \,,
\ee
manifestly depending only on $y$ and $\tilde y'$.

There is another proposal for an FM transform in the literature \cite{Demulder:2018lmj} which was motivated by the expected transformation of RR fields under PL T-duality \cite{Hassler:2017yza}; however, the precise relation of that to the present result remains unclear.

\paragraph{Proposed Fourier-Mukai kernel with spectators}

In terms of the objects introduced in section \ref{subsec:local}, we would have
\be
2 \varpi = 2 (\Phi - \widetilde\Phi') = (r^\nabla)\wedge (\tilde{l} + \widetilde{A}) -  (\tilde{r}'{}^\nabla)\wedge (l' + A') \,.
\ee
If $\mathbb{M}$ has bibundle structure satisfying the conditions of Proposition \ref{factorisation:prop}, we can rewrite this using  \eqref{def:caltildeAdef} in terms of the globally-defined (on $\mathbb M$) $\tG$-connection 1-form $\widetilde{\mathcal A}$  and the map $\tilde a:\mathbb{M}\to \inn \DD$ and their ``primed'' analogues (corresponding to the opposite parameterisation $\bbg=g'\tilde g'$ as opposed to the original $\bbg=\tilde g g$):
\be
2 \varpi   = (r^\nabla)\wedge \tilde a \widetilde{\mathcal A}-  (\tilde{r}'{}^\nabla)\wedge a' \mathcal A' \,.
\ee
This is globally well-defined, and of course reduces to the simpler spectator-free and abelian cases described above.
What is notable is that it involves both the genuine $G,\tG$ connections $\mathcal A$ and $\widetilde{\mathcal A}$ of \eqref{def:caltildeAdef} as well as the connection-like 1-forms $r^\nabla,\tilde r'{}^\nabla$ of \eqref{def:rnabla}.

\section{Discussion}
The main result of this paper has been to show how certain ``type 1'' bibundles (Definition \ref{def:type1}) provide examples of PL duality with spectators. Although we were able to exhibit topology change in the fibration structure (trivial vs.~non-trivial bundle), the type 1 restriction is a rather severe topological constraint that essentially excludes e.g.~simply connected spectator manifolds $B$ in the sense that they cannot accommodate non-trivial type 1 bibundles. It would be interesting to alleviate this  type restriction.  This leads to non-trivial bibundle structures even without spectators \cite{murray2012existence}, with underlying manifold $\DD$, right action the  canonical one of $\DD$ on itself, and a left action twisted by an outer automorphism.   It is likely   that pushing forward in this ``bibundle'' direction might elucidate some puzzles that remain even in the type 1 case that we treated, such as the geometric character (2-connection \cite{Aschieri:2003mw}?) of the form $r^\nabla$ of subsection \eqref{subsec:connfromproj} that arose in the reduction of the (ordinary) principal $\DD$-bundle connection $\bbalpha$.

We also obtained a formula for the Fourier-Mukai kernel that enters into the transformation of RR fluxes, that automatically respects the Bianchi identities in the sense that RR polyforms with $d\pm \mathsf{H}\wedge=0$ are mapped into dual ones that have $d\pm \widetilde{\mathsf{H}}\wedge=0$.   In the abelian case, one can prove that the FM transform descends to an isomorphism in cohomology. In the PL case that we are instead interested in, this statement will need to be qualified on account of how at least one of $G,\tG$ are non-compact in known examples. Our proposal for the FM transform agrees with that of \cite{Gevorgyan:2013xka} in the non-abelian case (where one group is abelian), but the relation with the FM transform proposed in \cite{Demulder:2018lmj} is currently unclear.

One virtue of the  the QP-manifold approach that we have taken, beyond its expediency for calculation,  is that it is amenable to generalisation to D-branes and M-theory  \cite{Arvanitakis:2018cyo,Chatzistavrakidis:2019seu,Arvanitakis:2021wkt}. This provokes the exciting notion of studying topological U-duality, including the recently proposed Poisson-Lie U-duality \cite{Sakatani:2019zrs,Malek:2019xrf} where the Drinfel'd double Lie algebra is replaced by a Leibniz algebra known as the exceptional Drinfel'd algebra. We hope to report on progress in these directions in the future.

\section*{Acknowledgements}

ASA, CB and DCT all acknowledge the support of the FWO-Vlaanderen through the project G006119N, as well as that of the Vrije Universiteit Brussel through the Strategic Research Program ``High-Energy Physics''. 
ASA and CB are also supported by FWO-Vlaanderen Postdoctoral Fellowships.
DCT is supported by The Royal Society through a University Research Fellowship {\em Generalised Dualities in String Theory and Holography} URF 150185 and in part by STFC grant ST/P00055X/1.  
We would like to thank Jeffery Giansiracusa and Fridrich Valach for helpful discussions, and Athanasios Chatzistavrakidis, Saskia Demulder and Richard Szabo for feedback on a draft. 

\appendix
\section{Notation and conventions}
\label{notcon}
\subsection{QP-manifolds}
A QP-manifold of degree $n$ is a graded manifold ${\cal M}$ equipped with a symplectic structure $\omega$, and a compatible cohomological vector field $Q$, a vector field of degree +1 that obeys $Q^2=0$ and $L_Q \omega =0$.   On the ring of functions $C^\infty(\cal M)$ $\omega$ induces a graded Poisson bracket $\{\bullet, \bullet\}$ obeying
\be
\label{PBidentities}
\begin{aligned}
\{f , g h \} &= \{f ,g \} h + (-1)^{(|f|-n)|g|}g \{f, h\}\,,\\
\{f, g \} &= - (-1)^{(|f|-n)(|g|-n)}\{g ,f\} \,,\\
 \{ f , \{g, h\} \}   &=  
\{ \{ f, g \} , h \}
+ (-1)^{(|f|-n)(|g|-n)} \{ g , \{ f,h\}\}\,.
\end{aligned}
\ee
Instead of working with $Q$ directly we will work with its generator $\Theta \in C^\infty(\cal M)$ of degree $n+1$,  called the Hamiltonian, such that 
\be
Q = \{\Theta, \bullet\} \, , \quad \{\Theta, \Theta\} = 0 \, . 
\ee
The second of these, the classical master equation, enforces nilpotency of $Q$. 

\subsection{General group and bundle conventions} 

The constructions described in this paper use the notation of tables \ref{tab:gpcons}, \ref{tab:buncons} and \ref{tab:qpcons} below. 

\begin{table}[h]
\renewcommand{\arraystretch}{1.25}
\begin{minipage}{.5\linewidth}
	\begin{tabular}{ |c||c|c|c| } 
	\hline
	Group & $\mathbb{D}$ &  $G$ &  $\widetilde{G}$  \\ 
	\hline 
	Element & $\mathbbm{g}$ & $g$ & $\widetilde{g}$\\ 
	Algebra & $\frak{d}$ & $\frak{g}$ & $\widetilde{\frak{g}}$  \\ 
	Generators  & $\mathbb{T}_A$ & $T_a$ & $\widetilde{T}^a$  \\  
	Structure & $\mathbb{F}_{AB}{}^C$ & $f_{ab}{}^c$  & $\widetilde{f}^{ab}{}_c$\\ 
	Left MC form  & $\mathbb{L}^A$ & $l^a$ & $\widetilde{l}_a$ \\ 
	Right MC form  & $\mathbb{R}^A$ & $r^a$ & $\widetilde{r}_a$ \\ 
	Left action  & $\mathbb{U}_A$ & $u_a$ & $\widetilde{u}^a$  \\ 
	Right action & $\mathbb{V}_A$ &  $v_a$  & $\widetilde{v}_a$\\
	Adjoint action & $\mathbbm{a}[\mathbbm{g}]_A{}^B$ &  $a[g]_a{}^b$   & $\widetilde{a}[\widetilde{g}]^a{}_b$\\
	Inner product & $\eta    $&  $\kappa$   & $\widetilde{\kappa}$\\
	\hline
	\end{tabular}
	\caption{Group and algebra conventions \label{tab:gpcons}}

\end{minipage}
\begin{minipage}{.5\linewidth}\centering
	\begin{tabular}{ |c||c|c|c| } 
	\hline
	Bundle & $\mathbb{M}$ &  $M$ &  $\widetilde M$  \\ 
	\hline 
	Global connection 1-form & $\bbalpha$ & $\mathcal A$ & $\widetilde{\mathcal A}$\\
	Global field strength 2-form & 
	${\digamma}$ & $\mathcal F$ & $\widetilde{\mathcal F}$\\
	Gauge field 1-form & $\mathbb{A}$ & $A$ & $\widetilde A$ \\
	Gauge field strength 2-form & $\mathbb{F}$ & $F$ & $\widetilde F$\\ \hline
	\end{tabular}
	\caption{Bundle conventions \label{tab:buncons}}
	\vspace{2em}
	
	\begin{tabular}{ |c||c|c|c| } 
	\hline
	Group & $\mathbb{D}$ &  $G$ &  $\widetilde{G}$  \\ 
	\hline 
	(deg 0) Coordinates & $\mathbb{Y}^I $ & $y^i$ & $\widetilde{y}_i$ \\ 
	(deg 1) Coordinates & $\bbtheta^A  $, $\bbxi_A  $ & $\theta^a$, $\xi_a$ & $\widetilde{\theta}_a$, $\widetilde{\xi}^a$ \\ 
	(deg 2) Coordinates & $\mathbbm{p}_I  $  & $p_i$  & $\widetilde{p}^i$  \\
	\hline
	\end{tabular}
	\caption{QP-manifold coordinates \label{tab:qpcons}}
\end{minipage}
\end{table}

\paragraph{Group and algebra conventions}  For a Lie algebra $\frak{d}$  corresponding to Lie groups $\mathbb{D}$, we let $\{\mathbb{T}_A\}$ be a basis  such that  $[\mathbb{T}_A, \mathbb{T}_B]=\mathbb{F}_{AB}{}^C \mathbb{T}_C$.  We define the left and right invariant Maurer-Cartan one-forms as 
\be
\mathbb{L} = -\mathbbm{g}^{-1}  d \mathbbm{ g}  =  \mathbb{L}^A  \mathbb{T}_A =  \mathbb{L}^{A}{}_I d\mathbb{Y}^I \mathbb{T}_A \, , \quad
\mathbb{R} = -  d \mathbbm{g}\mathbbm{g}^{-1}   =  \mathbb{R}^A  \mathbb{T}_A =  \mathbb{R}^{A}{}_I d \mathbb{Y}^I \mathbb{T}_A  \, , 
\ee 
which obey
\be
d \mathbb{L} =  \mathbb{L}\wedge \mathbb{L} \Rightarrow d \LL^A = \tfrac{1}{2} \mathbb{F}_{BC}{}^A \LL^B \wedge \LL^C \,, \quad d \mathbb{R} =  - \mathbb{R}\wedge \mathbb{R} \Rightarrow d \RR^A = -\tfrac{1}{2} \mathbb{F}_{BC}{}^A \RR^B \wedge \RR^C \,, 
\ee
The vector fields $\mathbb{V}_A = \mathbb{V}_{A}{}^I \partial_I$ dual to the left-invariant forms, $\iota_{\mathbb{V}_A} \mathbb{L}^B = \delta_A^B$, generate right action on $\mathbb{D}$ and obey $[\mathbb{V}_A, \mathbb{V}_B] = -\mathbb{F}_{AB}{}^C \mathbb{V}_C $.  These vectors leave invariant the right-invariant forms, $L_{\mathbb{V}_A} \mathbb{R} = 0$.
Similarly, the vectors fields $\mathbb{U}_A$ dual to the right-invariant forms, $\iota_{\mathbb{U}_A} \RR^B = \delta_A^B$, generate left action on $\mathbb{D}$ and obey $[\mathbb{U}_A, \mathbb{U}_B] = + \mathbb{F}_{AB}{}^C \mathbb{U}_C $ and $L_{\mathbb{U}_A} \mathbb{L} = 0$.

We let the adjoint action be $\mathbbm{g}^{-1} T_A \mathbbm{ g}  = \mathbbm{a}[ \mathbbm{ g} ]_A{}^BT_B$.
We will often leave the dependence on $\mathbbm{g}$ implict and omit the square brackets.
We have $d \mathbbm{a}_{A}{}^B = \mathbb{L}^C \mathbbm{a}_A{}^D \mathbb{F}_{CD}{}^B = \RR^C  \mathbbm{a}_{D}{}^B\mathbb{F}_{CA}{}^D$.
One has $\RR^A = (\mathbbm{a}^{-1})_B{}^A \LL^B$, $\LL^A = \mathbbm{a}_B{}^A \RR^B$ and hence $\iota_{\UU_A} \LL^B = \mathbbm{a}_A{}^B$, $\iota_{\VV_A} \RR^B = (\mathbbm{a}^{-1})_A{}^B$.
We denote the adjoint invariant inner product on $\mathfrak{d}$ as $\eta_{AB} = \langle  \mathbb{T}_A ,  \mathbb{T}_B \rangle $.  

Similar definitions hold for groups $G$ and $\widetilde{G}$ with the dictionary provided by table \ref{tab:gpcons}

\paragraph{Bundle conventions} 

We work with connections for principal left bundles (i.e. there is a global left action but transition functions use local right action). We define
\be
\bbalpha\equiv-d\mathbbm{g}\mathbbm{g}\inv + \mathbbm{g}\mathbb{A}\mathbbm{g}\inv \, , \qquad \digamma = d \bbalpha + \bbalpha \wedge \bbalpha = \mathbbm{g} \mathbb{F}\mathbbm{g}^{-1}\,,\quad
\FF = d \AA + \AA \wedge \AA \,,
\ee
such that
\be
\FF^A = d\AA^A + \tfrac{1}{2} \mathbb{F}_{BC}{}^A \AA^B \wedge \AA^C\,,\quad
\digamma^A = (a^{-1})_B{}^A \FF^B \,.
\ee
Under a local right transformation, $\bbalpha$ and $\digamma$ are invariant while 
\be
\mathbbm{g} \mapsto \mathbbm{g} \hh_R\,,\quad 
\AA\mapsto \hh_R^{-1} d \hh_R + \hh_R^{-1} \AA \hh_R \,,\quad
\mathbb{F} \mapsto \hh_R^{-1} \mathbb{F} \hh_R 
\ee
The Chern-Simons three-form is
\be\label{eq:CSlaw}
\begin{aligned}
\omega_{CS}(\bbalpha ) &= \langle \bbalpha  \, \overset{\wedge}{,}\, d \bbalpha  + \tfrac{2}{3} \bbalpha\wedge \bbalpha \rangle  = \bbalpha^A\wedge d\bbalpha^B\eta_{AB}+ \tfrac{1}{3}\bbalpha^A \wedge\bbalpha^B\wedge\bbalpha^C \eta_{AD} \mathbb{F}_{BC}{}{}^D \\
&=  \omega_{CS}(\mathbb{A} ) - \tfrac{1}{3}  \langle \mathbb{L}  \, \overset{\wedge}{,}\, \mathbb{L} \wedge  \mathbb{L}  \rangle - \langle \mathbb{A}  \, \overset{\wedge}{,}\, \mathbb{L} \wedge  \mathbb{L}  \rangle+ \langle \mathbb{L}  \, \overset{\wedge}{,}\, d \mathbb{A}  \rangle \\
&= \omega_{CS}(\mathbb{A} ) - \tfrac{1}{6} \mathbb{F}_{ABC} \mathbb{L}^A \wedge   \mathbb{L}^B \wedge  \mathbb{L}^C  -\tfrac{1}{2} \mathbb{F}_{ABC}   \mathbb{L}^A \wedge   \mathbb{L}^B \wedge  \mathbb{A }^C  +  \mathbb{L}^A \wedge  d\mathbb{A}_A \, .  
\end{aligned} 
\ee
The three-form $\omega_{CS}(\bbalpha )$ is a globally-defined section of $\Lambda^3 T^\star\mathbb M$; $ \omega_{CS}(\mathbb{A} )$ is the conventional Chern-Simons three-form, locally defined on $B$.

\subsection{Poisson-Lie conventions} 

\paragraph{Poisson-Lie group}
A Lie group $G$ is said to be Poisson-Lie (PL) when it is further equipped with a $\mathfrak{g}\wedge \mathfrak{g}$ valued one-cocycle $\Pi = \Pi^{ab}T_a \otimes T_b$ that  vanishes at the identity, is skew, compatible with right multiplication, and obeys in particular 
\begin{equation}\label{eq:dPi}
d\Pi^{ab} = - l^c \widetilde{f}^{ab}{}_c  - 2 l^c f_{cd}{}^{[a} \Pi^{b]d} \, . 
\end{equation} 
Moreover $\Pi^{ab}$ defines a Poisson structure on $G$.   Here $\widetilde{f}^{ab}{}_c$ are can be viewed as structure constants for an algebra $\widetilde{\mathfrak{g}}$, or as defining a $\mathfrak{g}\wedge \mathfrak{g}$ valued one-cocycle, $\delta$, on $\mathfrak{g}$ via $\delta(T_a) = \widetilde{f}^{bc}{}_a T_b \otimes T_c$.   In this way $(\mathfrak{g}, \delta)$ forms a Lie bi-algebra.

\paragraph{Drinfel'd double}
A Drinfel'd Double algebra $\mathfrak{d}$ is an even-dimensional Lie algebra equipped with an ad-invariant split sign-signature pairing $\eta$, that admits at least one decomposition $\mathfrak{d}= \mathfrak{g} + \widetilde{\mathfrak{g}} $ in which $\mathfrak{g}$ and $\widetilde{\mathfrak{g}} $ are sub-algebras maximally isotropic with respect to $\eta$.  The triple of algebras $\{ \frak{d}, \frak{g}, \widetilde{\mathfrak{g}}\}$ is called a Manin triple.  We denote the generators of each of the algebras contained in a  Manin triple as  $\mathbb{T}_A$, $T_a$, $\widetilde{T}^a$ respectively.  We then have that 
\be
\eta(T_a , T_b) = \eta(\widetilde{T}^a ,\widetilde{T}^b)=  0 \, ,  \, \quad \eta(T_a ,\widetilde{T}^b)= \eta( \widetilde{T}^b, T_a) =\delta^b_a \, . 
\ee
Adjoint invariance determines that 
\be
[T_a, T_b] = f_{ab}{}^c T_c \, , \quad [T_a, \widetilde{T}^b] = \widetilde{f}^{bc}{}_a T_c - f_{ac}{}^b \widetilde{T}^c  \, , \quad  [\widetilde{T}^a ,\widetilde{T}^b] = \widetilde{f}^{ab}{}_c \widetilde{T}^c \, . 
\ee
We can specify the adjoint action of $G= \exp \frak{g}$ via
\begin{align}
g^{-1} T_a g   = a_a{}^b[g] T_b\,,\quad
g^{-1} \widetilde T^a g  = b^{ab}[g] T_b + (a^{-1})_b{}^a[g] \widetilde T^b \,,
\end{align}
and we define the Poisson-Lie bivector of $G$ via
\be
\Pi^{ab}[g] \equiv b^{ca}[g] a_c{}^b[g]\,,
\ee
so $b^{ab} = (a^{-1})_c{}^a \Pi^{bc}$.
Acting with $gh$ rather than $g$ in the above expressions we obtain the composition laws
\be
a_a{}^b[gh] = a_a{}^c[g] a_c{}^b[h]\,,\quad
\Pi^{ab}[gh] = \Pi^{ab}[h] + a_c{}^a[h] \Pi^{cd}[g] a_d{}^b[h] \,.
\ee
We also have $\Pi^{ab}[e]=0$ by definition and 
\be
\Pi^{ab}[g^{-1}] = - (a^{-1})_c{}^a[g] b^{bc} [g] 
= - (a^{-1})_c{}^a [g]\Pi^{cd} [g] (a^{-1})_d{}^b[g]
\,.
\ee
Adjoint invariance of the structure constants implies
\be
f_{ef}{}^d a_d{}^c (a^{-1})_a{}^e (a^{-1})_b{}^f = f_{ab}{}^c 
\,,\quad 
\widetilde f^{ef}{}_d (a^{-1})_a{}^d a_e{}^b a_f{}^c 
= \widetilde f^{bc}{}_a + 2 f_{ad}{}^{[b} \Pi^{c]d} \,,
\label{adjinv2}
\ee
\be
2 \Pi^{c[a} \widetilde f^{b] e}{}_c + \Pi^{ac} \Pi^{bd} \widetilde f_{cd}{}^e - 2 \Pi^{c[a} f_{cd}{}^{b]} \Pi^{de} = -\widetilde f^{ab}{}_c \Pi^{ce} \,.
\label{eq:tricky}
\ee
The left-invariant forms $l^a$ and dual vector fields $v_a$ obey as above the differential identities $L_{v_a} v_b = - f_{ab}{}^c v_c$, $d l^a = + \frac{1}{2} f_{bc}{}^a l^b \wedge l^c$, $L_{v_a} l^b = f_{ac}{}^b l^c$.
In addition $\Pi^{ab}$ obeys \eqref{eq:dPi}. 
It is useful to also define $\pi^a \equiv \Pi^{ab} v_b$, such that
\be
L_{v_a} \pi^b = f_{ac}{}^b \pi^c - \widetilde f^{bc}{}_a v_c\,,\quad
L_{\pi^a} \pi^b = - \widetilde f^{ab}{}_c \pi^c \,.
\ee

We could also repeat the above identities starting with the adjoint action of $\widetilde G = \exp \widetilde{\frak{g}}$ on $\frak{d}$
\begin{align}
\tilde g^{-1} T_a \tilde g    = \widetilde b_{ab}[\tilde g] \widetilde T^b + (\widetilde a^{-1})^b{}_a[\tilde g] T_b\,, \quad
\tilde g^{-1} \widetilde T^a \tilde g & = \widetilde a{}^a{}_b[\tilde g] \widetilde T^b \,,
\end{align}
and defining the Poisson-Lie bivector of $\tG$ via
\be
\widetilde \Pi_{ab}[\tilde g] \equiv \widetilde b_{ca}[\tilde g] \widetilde a^c{}_b [\tilde g]\,.
\ee
This satisfies
\be
\label{eq:dPitilde}
d\widetilde \Pi_{ab}=- \tilde l_c f_{ab}{}{}^c-2 \tilde l_c\tilde f^{cd}{}{}_{[a} \widetilde \Pi_{b]d}\,.
\ee
Note that $\widetilde \Pi^\mathsf{L}[\tilde g]\equiv-\widetilde \Pi[\tilde g\inv]$ is relevant for the symplectic reduction of section \ref{condSympl}. That satisfies  
\be
\label{eq:dPitildeR}
d\widetilde \Pi^\mathsf{L}_{ab}=- \tilde r_c f_{ab}{}{}^c+2 \tilde r_c\tilde f^{cd}{}{}_{[a} \widetilde \Pi^\mathsf{L}_{b]d}\,.
\ee

\subsection{Parametrisation of Drinfel'd Double}
\label{paramdd}

 Locally, at least, we can always parameterise $\mathbbm{ g}\in \mathbb{D}$ in terms of group elements $g\in G$ and $\tilde{g}\in \tilde{G}$ by
\be
\label{doubleparameterisation}
 \mathbbm{g}(\mathbb{Y}) = \tilde{g} g  \, . 
 \ee
 A short calculation shows that the invariant one-forms decompose as 
 \be
 \mathbb{L} = g^{-1} \tilde{l} g + l \, \quad  \mathbb{R} = \tilde{r} + \tilde{g} r \tilde{g}^{-1} \, ,
 \ee
 whose  dual vector fields have components
 \be
 \mathbb{U}^a = \tilde{u}^a \, , \quad \mathbb{U}_{a} = \tilde{  a}[\tilde{g}^{-1}]^b{}_a u_b + \tilde{\Pi}[\tilde{g}^{-1}]_{ab} \tilde{u}^b  \, , 
 \ee
 and
\be
\mathbb{V}^a = \Pi[g]^{ab} v_b +  a[g]_b{}^a \tilde{v}^b  \, , \quad  \mathbb{V}_a = v_a \, . 
\ee

Globally speaking we only consider what we will call \emph{perfect} \drf{} doubles; namely groups $\DD$ for which (besides the conditions involving the Lie algebra)
\begin{enumerate}
\item there exists a diffeomorphism $\varphi: \tG\times G \to \DD$ from the product manifold $\tG\times G$ onto $\DD$ that has the form
\be
\label{originalfactorisation}
\varphi= \mathsf{m}_\DD(\tilde \iota \times \iota)
\ee
for $\iota,\tilde \iota$ 1--1 group homomorphisms $\iota: G\to \DD$ and $\tilde \iota:\tG\to \DD$ and $\mathsf{m}_\DD$ the multiplication in $\DD$;
\item $\iota G\cap \tilde \iota \tG=1\in \DD$.
\end{enumerate}
The map $\varphi$ is \textbf{not} a group homomorphism, unless $\DD$ is abelian. (These are called ``double Lie groups'' in \cite{lu1990poisson}, without the second condition.) This map $\varphi$ realises the parameterisation of \eqref{doubleparameterisation} globally. Examples include $\DD=T^\star G\cong G\times \mathfrak{g}^\star$ and $\DD=\mathrm{SL}(2;\mathbb{C})$ \cite{alekseevsky1998poisson}, both discussed in subsection \ref{subsec:examples}.

Given $\varphi$ we can also define a ``primed'' factorisation where $G$ and $\tG$ appear in the other order. This is explicitly given by the diffeomorphism $\varphi':G\times \tG\to \DD$ defined via its inverse (where $\mathsf{I}_G$ is the inversion map $g\to g^{-1}$ of $G$, resp.~for other groups, and $\sigma$ permutes $\tG\times G\to G\times \tG$):
\be
{\varphi'}^{-1}\equiv (\mathsf{I}_{G}\times \mathsf{I}_{\tG})\sigma \varphi^{-1}\mathsf I_{\DD}\,.
\ee
This is indeed the inverse of  
\be
\label{primedfactorisation}
\varphi'=\mathsf{m}_{\DD}(\iota\times \tilde\iota):G\times \tG\to\DD
\ee (note the order!):
\be
\begin{split}
{\varphi'}^{-1} \varphi'(g,\tilde g)={\varphi'}^{-1}(\iota g\tilde\iota \tilde g)= (\mathsf{I}_G\times \mathsf{I}_{\tG})\sigma\varphi\inv \Big((\iota g \tilde\iota\tilde g)\inv\Big)\\
=(\mathsf{I}_G\times \mathsf{I}_{\tG})\sigma\varphi\inv \Big(\tilde\iota \tilde g\inv \iota g\inv\Big)=(\mathsf{I}_G\times \mathsf{I}_{\tG})\sigma(\tilde g\inv,g\inv)=(g,\tilde g)\,.
\end{split}
\ee

\bibliography{CurrentBib}

\providecommand{\href}[2]{#2}\begingroup\raggedright\begin{thebibliography}{10}

\bibitem{Bouwknegt:2003vb}
P.~Bouwknegt, J.~Evslin, and V.~Mathai, {\it {T duality: Topology change from H
  flux}},  {\em Commun. Math. Phys.} {\bf 249} (2004) 383--415,
  [\href{http://arxiv.org/abs/hep-th/0306062}{{\tt hep-th/0306062}}].

\bibitem{Bouwknegt:2003zg}
P.~Bouwknegt, K.~Hannabuss, and V.~Mathai, {\it {T duality for principal torus
  bundles}},  {\em JHEP} {\bf 03} (2004) 018,
  [\href{http://arxiv.org/abs/hep-th/0312284}{{\tt hep-th/0312284}}].

\bibitem{courant1990dirac}
T.~J. Courant, {\it Dirac manifolds},  {\em Transactions of the American
  Mathematical Society} {\bf 319} (1990), no.~2 631--661.

\bibitem{liu1997manin}
Z.-J. Liu, A.~Weinstein, and P.~Xu, {\it Manin triples for lie bialgebroids},
  {\em Journal of Differential Geometry} {\bf 45} (1997), no.~3 547--574.

\bibitem{Cavalcanti:2011wu}
G.~R. Cavalcanti and M.~Gualtieri, {\it {Generalized complex geometry and
  T-duality}},  in {\em {A Celebration of the Mathematical Legacy of Raoul Bott
  (CRM Proceedings \& Lecture Notes) American Mathematical Society (2010)
  341-366. ISBN: 0821847775}}, p.~0821847775, 2011.
\newblock \href{http://arxiv.org/abs/1106.1747}{{\tt arXiv:1106.1747}}.

\bibitem{Barmaz:2013yua}
Y.~Barmaz, {\it {T-duality through BV Morphisms and BV Pushforwards in
  Topological Field Theories}},  \href{http://arxiv.org/abs/1308.1913}{{\tt
  arXiv:1308.1913}}.

\bibitem{Heller:2016abk}
M.~A. Heller, N.~Ikeda, and S.~Watamura, {\it {Unified picture of non-geometric
  fluxes and T-duality in double field theory via graded symplectic
  manifolds}},  {\em JHEP} {\bf 02} (2017) 078,
  [\href{http://arxiv.org/abs/1611.08346}{{\tt arXiv:1611.08346}}].

\bibitem{Bessho:2015tkk}
T.~Bessho, M.~A. Heller, N.~Ikeda, and S.~Watamura, {\it {Topological
  Membranes, Current Algebras and H-flux - R-flux Duality based on Courant
  Algebroids}},  {\em JHEP} {\bf 04} (2016) 170,
  [\href{http://arxiv.org/abs/1511.03425}{{\tt arXiv:1511.03425}}].

\bibitem{Carow-Watamura:2018iau}
U.~Carow-Watamura, N.~Ikeda, T.~Kaneko, and S.~Watamura, {\it {DFT in
  supermanifold formulation and group manifold as background geometry}},  {\em
  JHEP} {\bf 04} (2019) 002, [\href{http://arxiv.org/abs/1812.03464}{{\tt
  arXiv:1812.03464}}].

\bibitem{Severa:2018pag}
P.~Severa and F.~Valach, {\it {Courant algebroids, Poisson-Lie T-duality, and
  type II supergravities}},  {\em Commun. Math. Phys.} {\bf 375} (2020), no.~1
  307--344, [\href{http://arxiv.org/abs/1810.07763}{{\tt arXiv:1810.07763}}].

\bibitem{delaOssa:1992vci}
X.~C. de~la Ossa and F.~Quevedo, {\it {Duality symmetries from nonAbelian
  isometries in string theory}},  {\em Nucl. Phys.} {\bf B403} (1993) 377--394,
  [\href{http://arxiv.org/abs/hep-th/9210021}{{\tt hep-th/9210021}}].

\bibitem{Klimcik:1995ux}
C.~Klimcik and P.~Severa, {\it {Dual nonAbelian duality and the Drinfeld
  double}},  {\em Phys. Lett.} {\bf B351} (1995) 455--462,
  [\href{http://arxiv.org/abs/hep-th/9502122}{{\tt hep-th/9502122}}].

\bibitem{Klimcik:1995dy}
C.~Klimcik and P.~Severa, {\it {Poisson-Lie T duality and loop groups of
  Drinfeld doubles}},  {\em Phys. Lett.} {\bf B372} (1996) 65--71,
  [\href{http://arxiv.org/abs/hep-th/9512040}{{\tt hep-th/9512040}}].

\bibitem{Sfetsos:1997pi}
K.~Sfetsos, {\it {Canonical equivalence of nonisometric sigma models and
  Poisson-Lie T duality}},  {\em Nucl. Phys. B} {\bf 517} (1998) 549--566,
  [\href{http://arxiv.org/abs/hep-th/9710163}{{\tt hep-th/9710163}}].

\bibitem{breen2007bitorseurs}
L.~Breen, {\it Bitorseurs et cohomologie non ab{\'e}lienne},  in {\em The
  Grothendieck Festschrift}, pp.~401--476.
\newblock Springer, 2007.

\bibitem{Aschieri:2003mw}
P.~Aschieri, L.~Cantini, and B.~Jurco, {\it {NonAbelian bundle gerbes, their
  differential geometry and gauge theory}},  {\em Commun. Math. Phys.} {\bf
  254} (2005) 367--400, [\href{http://arxiv.org/abs/hep-th/0312154}{{\tt
  hep-th/0312154}}].

\bibitem{murray2012existence}
M.~Murray, D.~M. Roberts, and D.~Stevenson, {\it On the existence of
  bibundles},  {\em Proceedings of the London Mathematical Society} {\bf 105}
  (2012), no.~6 1290--1314.

\bibitem{Roytenberg:2002nu}
D.~Roytenberg, {\it {On the structure of graded symplectic supermanifolds and
  Courant algebroids}},  in {\em {Workshop on Quantization, Deformations, and
  New Homological and Categorical Methods in Mathematical Physics}}, 3, 2002.
\newblock \href{http://arxiv.org/abs/math/0203110}{{\tt math/0203110}}.

\bibitem{severa2001some}
P.~{\v{S}}evera, {\it Some title containing the words" homotopy" and"
  symplectic", eg this one},  {\em arXiv preprint math/0105080} (2001).

\bibitem{Severa:1999ay}
P.~Severa, {\it {On geometry of non-Abelian duality}},  {\em Math. Phys. Stud.}
  {\bf 23} (2001) 217--226.

\bibitem{vsevera2015poisson}
P.~{\v{S}}evera, {\it Poisson-lie t-duality and courant algebroids},  {\em
  arXiv preprint arXiv:1502.04517} (2015).

\bibitem{Arvanitakis:2018cyo}
A.~S. Arvanitakis, {\it {Brane Wess-Zumino terms from AKSZ and exceptional
  generalised geometry as an $L_\infty$-algebroid}},  {\em Adv. Theor. Math.
  Phys.} {\bf 23} (2019), no.~5 1159--1213,
  [\href{http://arxiv.org/abs/1804.07303}{{\tt arXiv:1804.07303}}].

\bibitem{Arvanitakis:2019cxy}
A.~S. Arvanitakis, {\it {Generalising Courant algebroids to M-theory}},  {\em
  PoS} {\bf CORFU2018} (2019) 127, [\href{http://arxiv.org/abs/1904.12361}{{\tt
  arXiv:1904.12361}}].

\bibitem{Hori:1999me}
K.~Hori, {\it {D-branes, T duality, and index theory}},  {\em Adv. Theor. Math.
  Phys.} {\bf 3} (1999) 281--342,
  [\href{http://arxiv.org/abs/hep-th/9902102}{{\tt hep-th/9902102}}].

\bibitem{Svoboda:2020msh}
D.~Svoboda, {\em {Born Geometry}}.
\newblock PhD thesis, U. Waterloo (main), 2020.

\bibitem{Marotta:2018myj}
V.~E. Marotta and R.~J. Szabo, {\it {Para-Hermitian Geometry, Dualities and
  Generalized Flux Backgrounds}},  {\em Fortsch. Phys.} {\bf 67} (2019), no.~3
  1800093, [\href{http://arxiv.org/abs/1810.03953}{{\tt arXiv:1810.03953}}].

\bibitem{Marotta:2019eqc}
V.~E. Marotta and R.~J. Szabo, {\it {Born sigma-models for para-Hermitian
  manifolds and generalized T-duality}},  {\em Rev. Math. Phys.} {\bf 33}
  (2021), no.~09 2150031, [\href{http://arxiv.org/abs/1910.09997}{{\tt
  arXiv:1910.09997}}].

\bibitem{Alekseev:2004np}
A.~Alekseev and T.~Strobl, {\it {Current algebras and differential geometry}},
  {\em JHEP} {\bf 03} (2005) 035,
  [\href{http://arxiv.org/abs/hep-th/0410183}{{\tt hep-th/0410183}}].

\bibitem{Alexandrov:1995kv}
M.~Alexandrov, A.~Schwarz, O.~Zaboronsky, and M.~Kontsevich, {\it {The Geometry
  of the master equation and topological quantum field theory}},  {\em Int. J.
  Mod. Phys. A} {\bf 12} (1997) 1405--1429,
  [\href{http://arxiv.org/abs/hep-th/9502010}{{\tt hep-th/9502010}}].

\bibitem{Arvanitakis:2021wkt}
A.~S. Arvanitakis, {\it {Brane current algebras and generalised geometry from
  QP manifolds}: {Or, \textquotedblleft{}when they go high, we go
  low\textquotedblright{}}},  \href{http://arxiv.org/abs/2103.08608}{{\tt
  arXiv:2103.08608}}.

\bibitem{rcohenbundlebookunpublished}
R.~L. Cohen, {\em Bundles, Homotopy, and Manifolds}.
\newblock 20XX.
\newblock Available at \url{http://math.stanford.edu/~ralph/book.pdf}.

\bibitem{sniatycki1983reduction}
J.~{\'S}niatycki and A.~Weinstein, {\it Reduction and quantization for singular
  momentum mappings},  {\em Letters in mathematical physics} {\bf 7} (1983),
  no.~2 155--161.

\bibitem{cattaneo2013supergeometric}
A.~S. Cattaneo and M.~Zambon, {\it A supergeometric approach to poisson
  reduction},  {\em Communications in Mathematical Physics} {\bf 318} (2013),
  no.~3 675--716.

\bibitem{Reid-Edwards:2010plo}
R.~A. Reid-Edwards, {\it {Bi-Algebras, Generalised Geometry and T-Duality}},
  \href{http://arxiv.org/abs/1001.2479}{{\tt arXiv:1001.2479}}.

\bibitem{Grigoriev:2000zg}
M.~A. Grigoriev, A.~M. Semikhatov, and I.~Y. Tipunin, {\it {BRST formalism and
  zero locus reduction}},  {\em J. Math. Phys.} {\bf 42} (2001) 3315--3333,
  [\href{http://arxiv.org/abs/hep-th/0001081}{{\tt hep-th/0001081}}].

\bibitem{Voronov:2001qf}
T.~Voronov, {\it {Graded manifolds and drinfeld doubles for Lie bialgebroids}},
   \href{http://arxiv.org/abs/math/0105237}{{\tt math/0105237}}.

\bibitem{nomizu1955reduction}
K.~Nomizu, {\it Reduction theorem for connections and its application to the
  problem of isotropy and holonomy groups of a riemannian manifold},  {\em
  Nagoya Mathematical Journal} {\bf 9} (1955) 57--66.

\bibitem{Azcarraga:2011hqa}
J.~A.~d. Azc\'arraga and J.~M. Izquierdo, {\em {Lie Groups, Lie Algebras,
  Cohomology and some Applications in Physics}}.
\newblock Cambridge University Press, 4, 2011.

\bibitem{Freed:1991bn}
D.~S. Freed and F.~Quinn, {\it {Chern-Simons theory with finite gauge group}},
  {\em Commun. Math. Phys.} {\bf 156} (1993) 435--472,
  [\href{http://arxiv.org/abs/hep-th/9111004}{{\tt hep-th/9111004}}].

\bibitem{Tseytlin:1990nb}
A.~A. Tseytlin, {\it {Duality Symmetric Formulation of String World Sheet
  Dynamics}},  {\em Phys. Lett.} {\bf B242} (1990) 163--174.

\bibitem{Rocek:1991ps}
M.~Rocek and E.~P. Verlinde, {\it {Duality, quotients, and currents}},  {\em
  Nucl. Phys.} {\bf B373} (1992) 630--646,
  [\href{http://arxiv.org/abs/hep-th/9110053}{{\tt hep-th/9110053}}].

\bibitem{Gevorgyan:2013xka}
E.~Gevorgyan and G.~Sarkissian, {\it {Defects, Non-abelian T-duality, and the
  Fourier-Mukai transform of the Ramond-Ramond fields}},  {\em JHEP} {\bf 03}
  (2014) 035, [\href{http://arxiv.org/abs/1310.1264}{{\tt arXiv:1310.1264}}].

\bibitem{Demulder:2018lmj}
S.~Demulder, F.~Hassler, and D.~C. Thompson, {\it {Doubled aspects of
  generalised dualities and integrable deformations}},  {\em JHEP} {\bf 02}
  (2019) 189, [\href{http://arxiv.org/abs/1810.11446}{{\tt arXiv:1810.11446}}].

\bibitem{Hassler:2017yza}
F.~Hassler, {\it {Poisson-Lie T-Duality in Double Field Theory}},  {\em Phys.
  Lett. B} {\bf 807} (2020) 135455,
  [\href{http://arxiv.org/abs/1707.08624}{{\tt arXiv:1707.08624}}].

\bibitem{Chatzistavrakidis:2019seu}
A.~Chatzistavrakidis, L.~Jonke, D.~L\"ust, and R.~J. Szabo, {\it {Fluxes in
  Exceptional Field Theory and Threebrane Sigma-Models}},  {\em JHEP} {\bf 05}
  (2019) 055, [\href{http://arxiv.org/abs/1901.07775}{{\tt arXiv:1901.07775}}].

\bibitem{Sakatani:2019zrs}
Y.~Sakatani, {\it {U-duality extension of Drinfel'd double}},  {\em PTEP} {\bf
  2020} (2020), no.~2 023B08, [\href{http://arxiv.org/abs/1911.06320}{{\tt
  arXiv:1911.06320}}].

\bibitem{Malek:2019xrf}
E.~Malek and D.~C. Thompson, {\it {Poisson-Lie U-duality in Exceptional Field
  Theory}},  {\em JHEP} {\bf 04} (2020) 058,
  [\href{http://arxiv.org/abs/1911.07833}{{\tt arXiv:1911.07833}}].

\bibitem{lu1990poisson}
J.-H. Lu and A.~Weinstein, {\it Poisson lie groups, dressing transformations,
  and bruhat decompositions},  {\em Journal of Differential geometry} {\bf 31}
  (1990), no.~2 501--526.

\bibitem{alekseevsky1998poisson}
D.~Alekseevsky, J.~Grabowski, G.~Marmo, and P.~W. Michor, {\it Poisson
  structures on double lie groups},  {\em Journal of Geometry and Physics} {\bf
  26} (1998), no.~3-4 340--379.

\end{thebibliography}\endgroup

\end{document}